\documentclass[12pt,preprint]{aastex}

\slugcomment{Not to appear in Nonlearned J., 45.}

\shorttitle{Infrared Observations of Dust around V1280 Sco}
\shortauthors{Sakon et al.}

\begin{document}


\title{Concurrent Formation of Carbon and Silicate Dust in Nova V1280 Sco}


\author{Itsuki Sakon\altaffilmark{1}, Shigeyuki Sako\altaffilmark{2}, Takashi Onaka\altaffilmark{1}, Takaya Nozawa\altaffilmark{3}, Yuki Kimura\altaffilmark{4}, Takuya Fujiyoshi\altaffilmark{5}, Takashi Shimonishi\altaffilmark{6}, Fumihiko Usui\altaffilmark{1}, Hidenori Takahashi\altaffilmark{2}, Ryou Ohsawa\altaffilmark{2}, Akira Arai\altaffilmark{7}, Makoto Uemura\altaffilmark{8}, Takahiro Nagayama\altaffilmark{9,10}, Bon-Chul Koo\altaffilmark{11}, Takashi Kozasa\altaffilmark{12}}


\altaffiltext{1}{Department of Astronomy, Graduate Schools of Science, University of Tokyo, 7-3-1 Hongo, Bunkyo-ku, Tokyo 113-0033, Japan}
\altaffiltext{2}{Institute of Astronomy, University of Tokyo, 2-21-1 Ohsawa, Mitaka, Tokyo 181-0015, Japan}
\altaffiltext{3}{National Astronomical Observatory of Japan, Mitaka, Tokyo 181-8588, Japan}
\altaffiltext{4}{Institute of Low Temperature Science, Hokkaido University, Sapporo 060-0819, JAPAN}
\altaffiltext{5}{Subaru Telescope, National Astronomical Observatory of Japan, 650 North A'ohoku Place, Hilo, HI 96720, USA}
\altaffiltext{6}{Frontier Research Institute for Interdisciplinary Sciences, Tohoku University, Aramaki aza Aoba 6-3, Aoba-ku, Sendai 980-8578, Japan}
\altaffiltext{7}{Koyama Astronomical Observatory, Kyoto Sangyo University, Motoyama, Kamigamo, Kita-ku, Kyoto, 603-8555, Japan}
\altaffiltext{8}{Hiroshima Astrophysical Science Center, Hiroshima University, Kagamiyama 1-3-1, Higashi-Hiroshima 739-8526, Japan}
\altaffiltext{9}{Department of Physics and Astronomy, Graduate School of Science and Engineering, Kagoshima University, 1-21-35 Korimoto, Kagoshima 890-0065, Japan}
\altaffiltext{10}{Department of Astrophysics, Nagoya University, Furo-cho, Chikusa-ku, Nagoya, Aichi 464-8602, Japan}
\altaffiltext{11}{Department of Physics and Astronomy, Seoul National University , 1 Gwanak-ro, Gwanak-gu, Seoul 151-742, Korea}
\altaffiltext{12}{Department of Cosmosciences, Graduate School of Science, Hokkaido University, Sapporo 060-0810, Japan}


\begin{abstract}
We present infrared multi-epoch observations of the dust forming nova V1280 Sco over $\sim$2000 days from the outburst. The temporal evolution of the infrared spectral energy distributions at 1272, 1616 and 1947 days can be explained by the emissions produced by amorphous carbon dust of mass (6.6--8.7)$\times$10$^{-8}$M$_{\odot}$ with a representative grain size of 0.01$~\mu$m and astronomical silicate dust of mass (3.4--4.3)$\times$10$^{-7}$M$_{\odot}$ with a representative grain size of 0.3--0.5$~\mu$m. Both of these dust species travel farther away from the white dwarf without an apparent mass evolution throughout those later epochs. 
The dust formation scenario around V1280 Sco suggested from our analyses is that the amorphous carbon dust is formed in the nova ejecta followed by the formation of silicate dust either in the expanding nova ejecta or as a result of the interaction between the nova wind and the circumstellar medium.
\end{abstract}

\section{Introduction}
Understanding the formation process of circumstellar dust is a fundamental step in exploring the origin of interstellar dust. From an observational point of view, however, we have limited knowledge about how dust condenses in gaseous stellar ejecta containing nucleo-synthesized heavy elements and how the dusty circumstellar environment is consequently formed. Classical novae provide unique opportunities to observe those processes in a realistic timescale thanks to their relatively frequent occurrence in the nearby universe ($<$ a few kpc from the sun). The composition of dust formed in the stellar ejecta depends on the chemical properties of the stellar atmosphere; carbonaceous dust is formed when the C/O ratio in the envelope exceeds unity (carbon-rich), while silicate dust is formed when the C/O ratio is less than unity (oxygen-rich) \citep{wat04}. Recent infrared observations of some dusty novae, however, have detected the presence of emission from both carbonaceous and silicate dust: e.g., Nova V842 Cen \citep{smi94}; Nova QV Vul \citep{geh92}, Nova V705 Cas \citep{eva05}. The origin of such bimodal dust species, so far, has not been fully understood. Infrared monitoring observations of nearby dusty novae over several years from their outbursts are crucial in investigating the temporal evolution of the dust properties at different epochs and in uncovering the dust formation history in the circumstellar and interstellar environments around the novae.

V1280 Sco was discovered on 2007 Feb. 4.86 by Japanese amateur astronomers Y. Nakamura and Y. Sakurai \citep{yam07}. V1280 Sco is classified as a FeII nova from its early optical spectrum \citep{mun07} and, thus, is caused by an explosion on a CO white dwarf. The onset of dust formation around V1280 Sco was reported on 23 days after the discovery \citep{das07,rud07}. These characteristics are consistent with a classification of V1280 Sco as a CO nova \citep{das07}. The dust formation peak was reported between 36 and 45 days in the nova wind ejected predominantly on Day 10.5 \citep{che08} with a wind speed of 500 km s$^{-1}$ \citep{yam07}. An expanding dust shell moving at 0.35$\pm$0.03 mas day$^{-1}$ has been reported based on VLTI/AMBER and MIDI observations performed between Days 23 and 145 \citep{che08}. In addition to the major mass loss peak on Day 10.5, the rebrightening of V1280 Sco was observed on Day 110, which may correspond to the secondary minor mass loss peak \citep{das07}. The light curve evolution of V1280 Sco is extremely slow and, therefore, the mass of the white dwarf is estimated to be $M_{\rm{WD}}=0.6M_{\odot}$ or smaller \citep{hou10, nai12}. The distance to V1280 Sco is still under debate. A distance $D$ of 1.6 kpc is estimated by \citet{che08}, $D=1.25$ kpc by \citet{das08} and 0.63 kpc by \citet{hou10}. \citet{nai12} have recently revised the velocity of the dust shell to be 350$\pm$160 km s$^{-1}$ and estimated the distance to be $D=1.1\pm 0.5$ kpc. In the following analyses, we adopt $D=1.1$ kpc as the distance to dust around V1280 Sco.

This study investigates the temporal evolution of the infrared emission from the dusty nova V1280 Sco with the objectives of determining the temporal evolution of dust properties and examining how the dusty circumstellar environment is created around the nova on a timescale of several years following the outburst. Details of our observations are described in \S\ref{sect2}. The evolution of the mid-infrared images of V1280 Sco over 2000 days following the outburst is presented in \S\ref{sect3}. The near-infrared spectrum of V1280 Sco on Day 940 is shown in \S\ref{sect4}. The properties and the origin of dust around V1280 Sco are discussed  in \S\ref{sect5}. The dust and molecular band features in the near- and mid-infrared spectra of V1280 Sco are discussed in \S\ref{sect6}.

\section{Observations \label{sect2}}
N-band (8--13~$\mu$m) \& Q-band (17--26~$\mu$m) imaging and N-band low-resolution (NL) spectroscopic data of V1280 Sco on Day 150 were collected with the Cooled Mid-Infrared Camera and Spectrometer \citep[COMICS;][]{kat00} on the Subaru Telescope and those on Days 1272, 1616, and 1947 were collected with the Thermal-Region Camera Spectrograph \citep[TReCS;][]{deb05} on the Gemini-South telescope. In addition, a near-infrared (2.55--4.9~$\mu$m) spectrum on Day 940 was taken with the AKARI/Infrared Camera \citep[IRC;][]{ona07}. Long-term photometric monitoring observations in the optical and near-infrared have been made continuously from the very initial phase of the outburst using the Optical and Infrared Synergetic Telescopes for Education and Research (OISTER). A summary of the imaging observations is given in Table~\ref{tbl_obs_summary}.
\begin{deluxetable}{llccccc}
\tabletypesize{\scriptsize}
\tablecaption{Summary of the Imaging Observations \label{tbl_obs_summary}}
\tablehead{
\colhead{Epoch\tablenotemark{a}} & \colhead{Instrument} & \colhead{Band} & \colhead{Wavelength($\mu$m)} & \colhead{Band Width($/mu$m)} & \colhead{Int. Time} & \colhead{Date} }
\startdata
{\it{Day 150}} & Kanata/TRISPEC & J/Ks & 1.25/2.15 & 0.26/0.32 & 0.5s/0.1s & 26 Jun., 2007 \\
 & Subaru/COMICS & N8.8 & 8.8 & 0.8  & 20s & 7 Jul., 2007  \\
 & Subaru/COMICS & N11.7 & 11.7 & 1.0 & 20s & 7 Jul., 2007 \\
 & Subaru/COMICS & Q18.8 & 18.75 & 0.90 & 60s & 7 Jul., 2007  \\
 & Subaru/COMICS & Q24.5 & 24.56 & 0.75 & 60s & 7 Jul., 2007 \\
{\it{Day 1272}} & GAO/GIRCS & J/H/Ks & 1.25/1.65/2.15 & 0.26/0.29/0.32 & 360s & 26 AUG., 2010 \\
 & Gemini-S/TReCS & Si-1 & 7.73 & 0.69 & 120s & 1 AUG., 2010  \\
 & Gemini-S/TReCS & Si-3 & 9.69 & 0.93 & 120s & 1 AUG., 2010 \\
 & Gemini-S/TReCS & Si-5 & 11.66 & 1.13 & 120s & 1 AUG., 2010 \\
 & Gemini-S/TReCS & Qa & 18.30 & 1.51 & 120s & 1 AUG., 2010  \\
 & Gemini-S/TReCS & Qb & 24.56 & 1.92 & 600s & 1 AUG., 2010 \\
{\it{Day 1616}} & IRSF/SIRIUS & J/H/Ks & 1.22/1.65/2.16 & 0.26/0.29/0.32 & 64s & 27 JUL., 2011 \\
 & Gemini-S/TReCS & Si-1 & 7.73 & 0.69 & 60s & 10 JUL., 2011  \\
 & Gemini-S/TReCS & Si-3 & 9.69 & 0.93 & 60s & 10 JUL., 2011 \\
 & Gemini-S/TReCS & Si-5 & 11.66 & 1.13 & 60s & 10 JUL., 2011 \\
 & Gemini-S/TReCS & Qa & 18.30 & 1.51 & 120s & 10 JUL., 2011  \\
 & Gemini-S/TReCS & Qb & 24.56 & 1.92 & 120s & 10 JUL., 2011 \\
{\it{Day 1947}} & IRSF/SIRIUS & J/H/Ks & 1.22/1.65/2.16 & 0.26/0.29/0.32 & 480s & 26 SEP., 2012 \\
 & Gemini-S/TReCS & Si-1 & 7.73 & 0.69 & 60s & 6 JUN., 2012  \\
 & Gemini-S/TReCS & Si-3 & 9.69 & 0.93 & 60s & 6 JUN., 2012 \\
 & Gemini-S/TReCS & Si-5 & 11.66 & 1.13 & 60s & 6 JUN., 2012 \\
 & Gemini-S/TReCS & Qa & 18.30 & 1.51 & 120s & 6 JUN., 2012  \\
 & Gemini-S/TReCS & Qb & 24.56 & 1.92 & 120s & 6 JUN., 2012 \\
\enddata
\tablenotetext{a}{The epochs of the observations with Subaru/COMICS, Gemini-S/TReCS in days after the discovery.}
\end{deluxetable}

\begin{deluxetable}{llccccc}
\tabletypesize{\scriptsize}
\tablecaption{Summary of Spectroscopic Observations \label{t2}}
\tablehead{
\colhead{Epoch\tablenotemark{a}} & \colhead{Instrument} & \colhead{Mode} & \colhead{Wavelength ($\mu$m)} & \colhead{Resolution} & \colhead{Int. Time} & \colhead{Date} }
\startdata
{\it{Day 150}} & Subaru/COMICS & NL & 7.8--13.3 & $R\sim$160 near 10~$\mu$m  & 20s & 7 Jul., 2007 \\
{\it{Day 940}} & AKARI/IRC & NG & 2.55--4.9 & $R\sim100$ near 3~$\mu$m & 40s & 8 Sep., 2009 \\
{\it{Day 1272}} & Gemini-S/TReCS & NL & 7.70--12.97 & $R\sim100$ near 10~$\mu$m & 600s & 1 AUG., 2010 \\
{\it{Day 1616}} & Gemini-S/TReCS & NL & 7.70--12.97 & $R\sim100$ near 10~$\mu$m & 60s & 10 JUL., 2011 \\
{\it{Day 1947}} & Gemini-S/TReCS & NL & 7.70--12.97 & $R\sim100$ near 10~$\mu$m & 600s & 6 JUN., 2012 \\
\enddata
\tablenotetext{a}{The epochs of the observations with Subaru/COMICS, Gemini-S/TReCS and AKARI/IRC in days after the discovery.}
\end{deluxetable}
\pagebreak

\subsection{Subaru/COMICS data}
Imaging observations of V1280 Sco on Day 150 were made with four medium-band filters, N8.8, N11.7, Q18.8 and Q24.5. The center wavelength, $\lambda_c$, and band width, $\Delta \lambda$, are summarized in Table~\ref{tbl_obs_summary}. The plate scale of the detector was 0".13 and the field of view was 41" $\times$ 31". To cancel the high background radiation, the secondary mirror was chopped at a frequency of $\sim$0.5 Hz with a 10" throw in the north-south direction. The chopped image was subtracted from the one taken immediately before to eliminate sky emission. Since the chopping residual pattern was almost negligible for bright objects, nodding was not employed. Images collected at both chopping positions were used. The total on-source integration times are summarized in Table~\ref{tbl_obs_summary}. The data reduction was performed in the standard manner for mid-infrared ground-based imaging observations, including dark subtractions and flat fielding. The flux calibration of V1280 Sco was achieved using measurements of the two standard stars listed in \citet{coh99}; $\lambda$ Sgr for N8.8, N11.7, Q18.8, and $\mu$ Cep for Q24.5. The aperture size for the photometry was set to be 2" in radius.

For the NL spectroscopy of V1280 Sco on Day 150, a slit width of 0".33 was used. As in the case of the imaging observations, only the secondary mirror chopping was performed at a frequency of $\sim$0.5 Hz with a 10" throw in the north-south direction. The position angle of the slit was set to 0 degrees so that images at both chopping positions would fall on the slit. Spectra collected at both chopping positions were coadded resulting in the total on-source integration time of 20s for NL spectroscopy (see Table~\ref{t2}). The data reduction was performed in the standard manner for mid-infrared ground-based spectroscopic observations, including dark subtraction, flat fielding using domeflat, wavelength calibration using atmospheric lines, and distortion correction. Finally, a one-dimensional spectrum of V1280 Sco was extracted. Atmospheric absorption in the extracted spectrum was corrected by division by the spectrum of a standard star, $\lambda$ Sgr, whose absolute spectrum is provided in \citet{coh99}. The absolute flux calibration including the slit efficiency correction was performed by simply scaling the flux level of the obtained spectrum to match with the photometric data in the 8.8 and 11.7~$\mu$m bands.


\subsection{Gemini-S/TReCS data}
Mid-infrared imaging and spectroscopic observations of V1280 Sco on Days 1272, 1616, and 1947 were carried out at the Gemini-South observatory under the Subaru/Gemini Time Exchange Programmes GS-2010B-C-7 on 2010 August 1, GS-2011B-C-4 on 2011 July 10, and GS-2012A-C-5 on 2012 June 6. For the imaging portion, five medium-band filters, Si-1, Si-3, Si-5, Qa and Qb, were used. The center wavelength, $\lambda_c$, and band width, $\Delta \lambda$, are summarized in Table~\ref{tbl_obs_summary}. The pixel scale of the T-ReCS detector was 0".09 and the field of view was 28".8 $\times$ 21".6. Both chopping and nodding were performed to cancel the high background radiation and the chopping residual pattern. The chopping was performed at a frequency of $\sim$3 Hz with a 15" throw in the north-south direction. The on-source integration time is summarized in Table~\ref{tbl_obs_summary}. The {\it{mireduce}} task included in the Gemini IRAF package (https://www.gemini.edu/node/10795) was used to produce final coadded, chop-subtracted and nod-subtracted data from the raw data files. The flux calibration was accomplished by observing the Cohen standard star HD151680 \citep{coh99}. The aperture size for the photometry was set to be 2" in radius.

For the N-band low-resolution spectroscopy of V1280 Sco on Days 1272, 1616, and 1946, a slit width of 0".35 was used. As in the case of the imaging observations, both chopping and nodding were employed. The chopping was performed at a frequency of $\sim$2 Hz with a 15" throw in the north-south direction. The instrument position angle was set to 120 deg so that the slit would approximately run along the long axis of V1280 Sco. The on-source integration time is summarized in Table~\ref{t2}. The {\it{msreduce}} task included in the Gemini IRAF package was used to reduce the mid-infrared spectrum of V1280 Sco at each epoch. The wavelength calibration was carried out using telluric atmospheric lines. Finally, a one-dimensional spectrum of V1280 Sco was extracted. Atmospheric absorption in the extracted spectrum was corrected by division by the spectrum of a standard star, HD151680, whose calibrated spectrum is provided in \citet{coh99}. The absolute flux calibration including the slit efficiency correction was performed by simply scaling the flux of the obtained spectra to match the photometric data in the Si-1, Si-3 and Si-5 bands.

\subsection{AKARI/IRC data}
AKARI \citep{mur07} performed the warm mission after the exhaustion of liquid Helium on August 2007 until the termination of the mission on November 2011. A near-infrared (2.55--4.9~$\mu$m) spectrum of V1280 Sco on Day 940 was taken with AKARI/IRC as part of the AKARI phase 3-II Open Time Program '{\it{Spectral Evolution of Novae in the Near-IR Based on AKARI Observations (SENNA)}}' (PI.; I. Sakon). Pointing IDs of the data are 5201030.1 and 5201030.2. Each observation was performed with the IRCZ4 AKARI astronomical Observing Template (AOT) using the NIR grism disperser \citep[NG;][]{ohy07}. The target was placed in the Np aperture, a 1' $\times$ 1' field of view for the spectroscopy of a point source in the NIR channel. Since the image size of the IRC/NIR channel has a FWHM of $\sim$4".7, which is much larger than the size of V1280 Sco even at the later epochs (see \S~\ref{sect3}), V1280 Sco on Day 940 can be regarded as a point-source when observed with IRC on AKARI.

A single pointed observation performed with this AOT produces, sequentially, five dark frames, four spectra with NG, a reference image with N3 (3.2~$\mu$m), five spectra with NG and five dark frames. Each exposure consists of one short and one long exposures. Since the long-exposure images were saturated, we only used the short-exposure images in the present study. The final dark frame was produced by median-filtering ten short-exposure dark frames to correct for any cosmic-ray effects and it was then subtracted off from each short-exposure science frame.

The signal pattern originating from the sky background and/or foreground emission in the Np aperture was carefully removed. This pattern was reproduced by convolving the foreground and/or background spectra collected at the Ns slit over the whole area of the sky within the Np aperture assuming that the sky brightness is uniform \citep{sak12}. In order to remove the cosmic ray hits, we applied a 3-$\sigma$ clipping of 18 dark-subtracted short-exposed NG spectra. In this process, bad pixels identified in the dark image were masked out and a shift in position due to the pointing uncertainty during a pointed observation was estimated by the 0th-order images of bright sources in the FOV. The spectrum of V1280 Sco was extracted by using an extraction box with the width of 5 pixels (7".3) in spatial direction and the length of 300 pixels in dispersion direction. The extracted spectrum was divided by the NG system response curve to obtain the final flux-calibrated 2.5-5~$\mu$m spectrum of V1280 Sco on Day 940.

\subsection{Near-Infrared Photometric Data collected through OISTER}
Almost simultaneously with the Subaru/COMICS observations on Day 150 (7 July 2007), near-infrared photometric observations of V1280 Sco in the J and Ks bands were carried out on 26 June 2007 using the Triple Range Imager and Spectrograph \citep[TRISPEC;][]{wat05} on the 1.5-m {\it{KANATA}} telescope at Higashi-Hiroshima Observatory. Also almost simultaneous with the Gemini-S/TReCS observation on Day 1272 (1 August 2010), near-infrared photometric observations of V1280 Sco in the J, H, and Ks bands were carried out on 26 Aug 2009 using the Gunma Infrared Camera and Spectrograph \citep[GIRCS;][]{tak09} on the 1.5-m telescope at Gunma Astronomical Observatory \citep[GAO;][]{has02}. Finally, almost simultaneously with the Gemini-S/TReCS observations on Days 1616 (10 July 2011) and 1947 (6 June 2012), near-infrared photometric observations of V1280 Sco in the J, H, and Ks bands were carried out on 27 July 2011 and 27 Sep 2012, respectively, using Simultaneous three-color Infrared Imager for Unbiased Surveys \citep[SIRIUS;][]{nag01} mounted on the 1.4-m telescope of the IRSF (Infrared Survey Facility) at the South African Astronomical Observatory. The center wavelength, $\lambda_c$, the band width, $\Delta \lambda$, and the integration times for those observations are given in Tables~\ref{tbl_obs_summary}.


\section{Temporal Evolution of Mid-Infrared Images of V1280 Sco \label{sect3}}
\begin{figure}[htbp!]
\begin{center}
\includegraphics[width=0.7\linewidth]{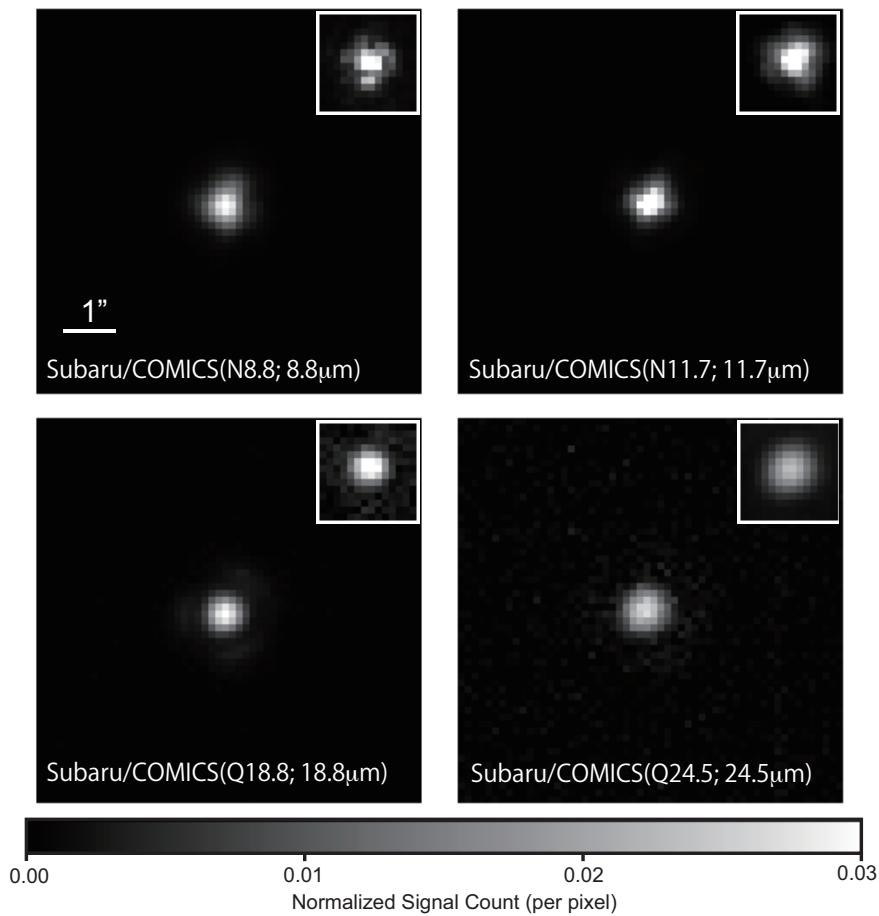}
\caption{The mid-infrared images of V1280 Sco on Day 150 taken at N8.8 (8.8~$\mu$m), N11.7 (11.7~$\mu$m), Q18.8 (18.8~$\mu$m) and Q24.5 (24.5~$\mu$m) using Subaru/COMICS. The image of a standard star ($\lambda$ Sgr for N8.8 and N11.7, $\mu$ Cep for Q18.8 and Q24.5) is shown in the right upper corner of each panel for comparison. The signal of each image is normalized by the total signals within 1" radius aperture. The 1" scale bar is shown in the first panel. North is to the left and east is toward the bottom in each image. \label{IMG0150}}
\end{center}
\end{figure}

\begin{figure}[htbp!]
\begin{center}
\includegraphics[width=\linewidth]{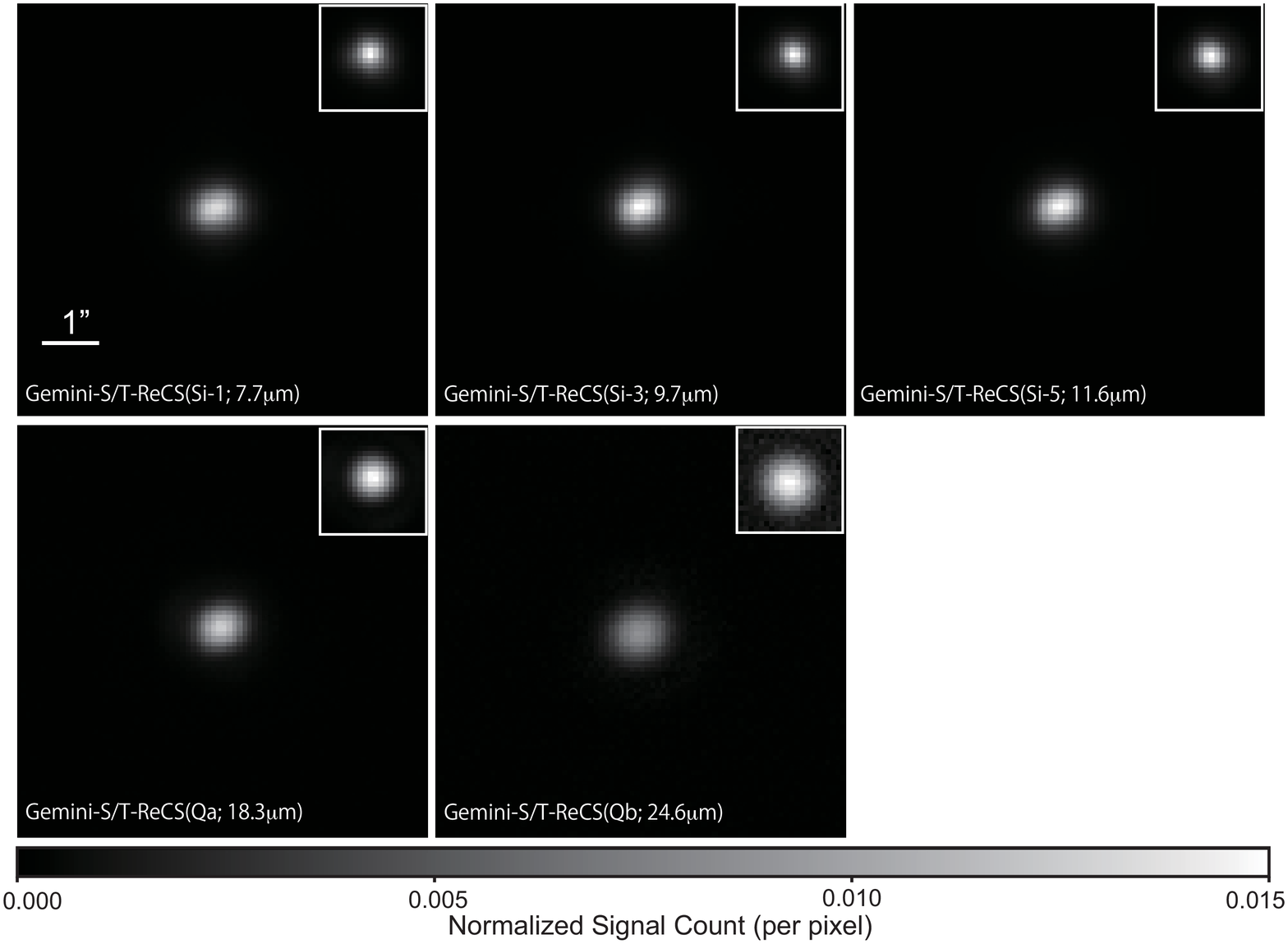}
\caption{The mid-infrared images of V1280 Sco on Day 1272 taken at Si-1 (7.73~$\mu$m), Si-3 (9.69~$\mu$m), Si-5 (11.66~$\mu$m), Qa (18.3~$\mu$m) and Qb (24.56~$\mu$m) using Gemini-S/TReCS. The image of standard star HD151680 is shown in the right upper corner of each panel for comparison. The signal of each image is normalized by the total signals within 1" radius aperture. The 1" scale bar is shown in the first panel. North is to the left and east is toward the bottom in each image.　\label{IMG1272}}
\end{center}
\end{figure}

\begin{figure}[htbp!]
\begin{center}
\includegraphics[width=\linewidth]{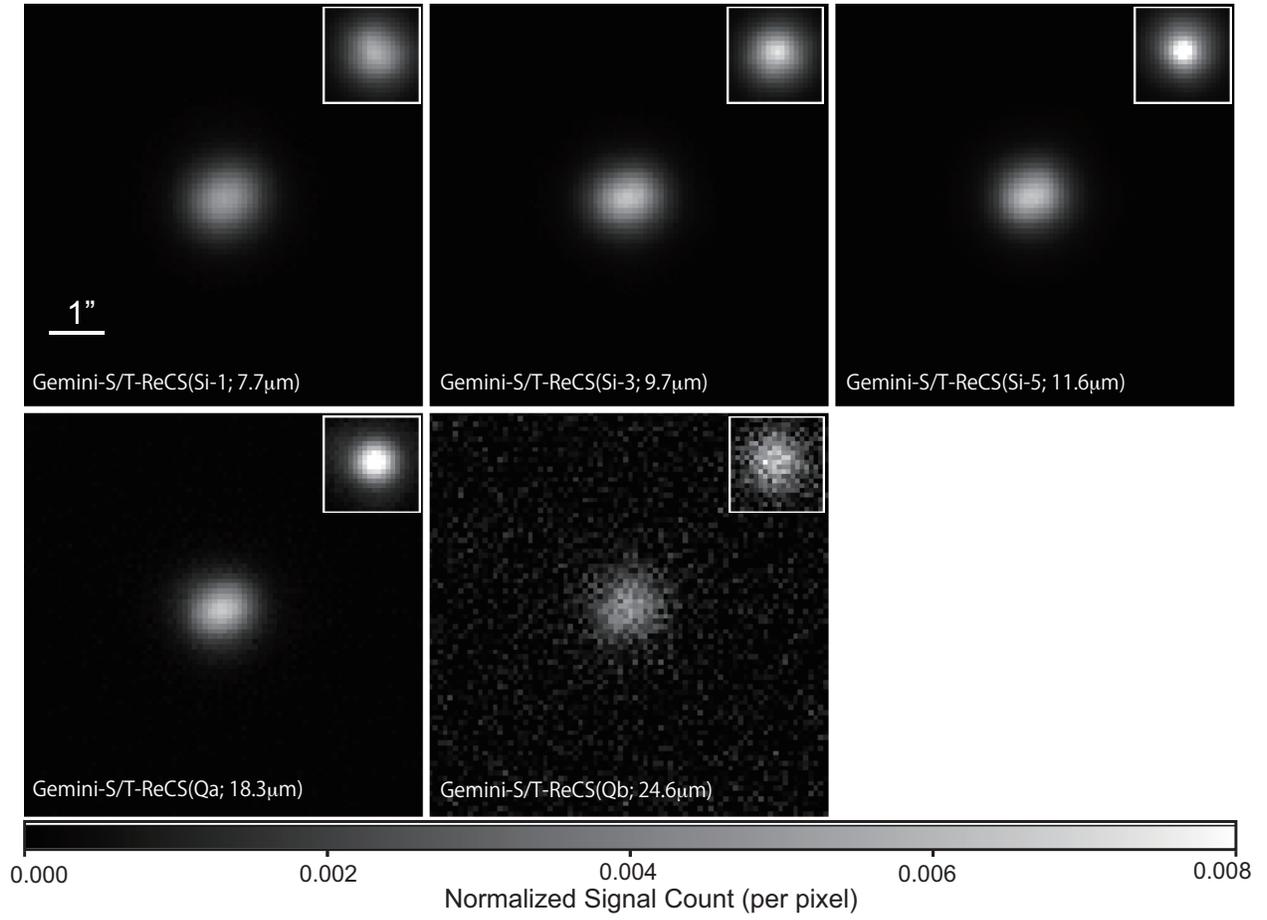}
\caption{The same as Fig.~\ref{IMG1272} but for the epoch of Day 1616.　\label{IMG1616}}
\end{center}
\end{figure}

\begin{figure}[htbp!]
\begin{center}
\includegraphics[width=\linewidth]{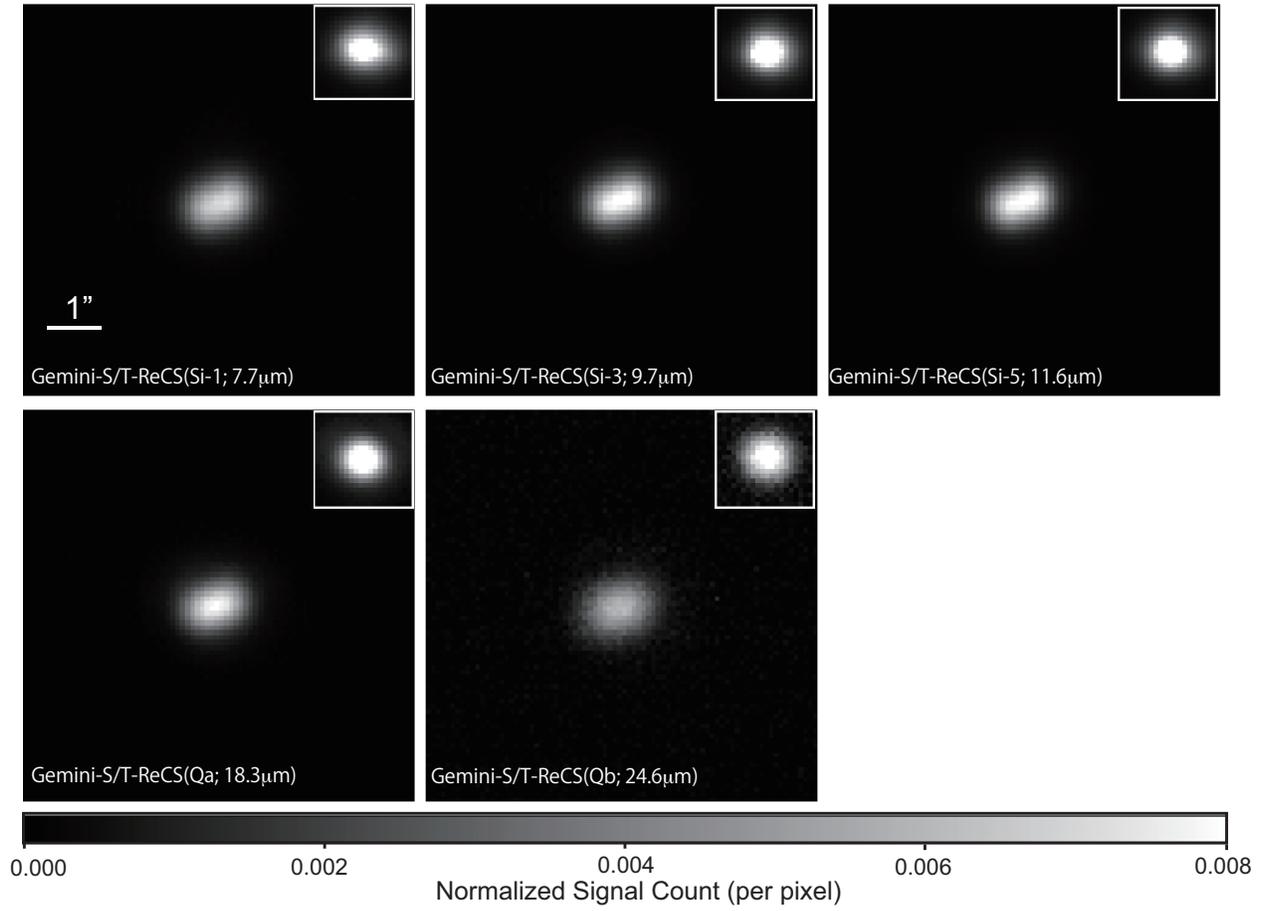}
\caption{The same as Fig.~\ref{IMG1272} but for the epoch of Day 1947. \label{IMG1947}}
\end{center}
\end{figure}

The mid-infrared images of V1280 Sco on Days 150, 1272, 1616 and 1947 are shown in Figures~\ref{IMG0150}--\ref{IMG1947}, respectively. On Day 150, the size of the emitting region cannot be spatially resolved with the Subaru/COMICS at any wavelengths. The intrinsic size of the emitting region of V1280 Sco at each wavelength $\lambda$ is modeled by a circular Gaussian profile given by
\begin{eqnarray}
G_{\lambda}(x,y)= V^{tot}_{\lambda} (2\pi \sigma_{\lambda}^2)^{-\frac{1}{2}}  exp\left\{ -\frac{1}{2} \left( \frac{(x-x_0)^2+(y-y_0)^2}{\sigma_{\lambda}^2} \right) \right\}, \label{eq_circ_gaussian}
\end{eqnarray}
where ($x_0$,$y_0$) are the central coordinates of the emitting region in pixels, $V^{tot}_{\lambda}$ is the flux density of V1280 Sco at wavelength $\lambda$, and $\sigma_{\lambda}$ is the size of the emitting region represented by the Gaussian dispersion in pixels. The parameters in Eq.~(\ref{eq_circ_gaussian}) are determined by deconvolving the image of V1280 Sco in each band with a point spread function of the standard star at the same wavelength. At all the wavelengths used, the obtained values of $\sigma_{\lambda}$ are within a range of the positional uncertainty due to the chopping stability in an exposure ($\sim 1$ pixel $=$0".13). Specifically, the size obtained for the emitting region at the longest wavelength, 24~$\mu$m, on Day 150 was $\sigma_{24.5\mu m}=0.5$ pixel, which corresponds to a distance of 0".065 from the white dwarf. Because of the large positional uncertainty ($\sim 1$ pixel) mentioned above, 0".065 is regarded as an upper limit. Near- and mid-infrared photometry of V1280 Sco on Day 150 is summarized in Table~\ref{tbl_photometry_150}.

\begin{deluxetable}{llcc}
\tabletypesize{\scriptsize}
\tablecaption{Results of the Near- to Mid-Infrared Photometory on Day 150. \label{tbl_photometry_150}}
\tablewidth{0pt}
\tablehead{
\colhead{Instrument} & \colhead{Band} & \colhead{Wavelength ($\mu$m)} & \colhead{Flux Density (Jy)}
 }
\startdata
KANATA/TRISPEC &  J   & 1.25 &   0.5$\pm$0.03 \\  
KANATA/TRISPEC &  Ks  & 2.15 &  10.6$\pm$0.20 \\
Subaru/COMICS & N8.8  &  8.8 & 131.5$\pm$2.0  \\
Subaru/COMICS & N11.7 & 11.7 & 125.5$\pm$2.0  \\
Subaru/COMICS & Q18.8 & 18.8 &  66.1$\pm$2.5  \\
Subaru/COMICS & Q24.5 & 24.5 &  24.5$\pm$2.5  \\
\enddata
\end{deluxetable}

On Days 1272, 1616 and 1947, all the mid-infrared images of V1280 Sco exhibit elongation along PA$=20$deg. In particular, the morphology on Day 1947 is very similar to that observed by \citet{che12} on Day 1589. Near- and mid-infrared photometric results for V1280 Sco on Days 1272, 1616 and 1947 are listed in Table~\ref{tbl_photometry_1272_1616_1947}. 

\begin{deluxetable}{lllcc}
\tabletypesize{\scriptsize}
\tablecaption{Results of the Near- to Mid-Infrared Photometry on Days 1272, 1616 and 1947. \label{tbl_photometry_1272_1616_1947}}
\tablewidth{0pt}
\tablehead{
\colhead{Epoch} & \colhead{Instrument} & \colhead{Band} & \colhead{Wavelength ($\mu$m)} & \colhead{Flux Density (Jy)} 
}
\startdata
{\it{Day 1272}}  & GAO/GIRCS & J & 1.24 & 0.635$\pm$0.013 \\  
& GAO/GIRCS & H & 1.66 & 0.683$\pm$0.017 \\
& GAO/GIRCS & Ks & 2.16 & 0.824$\pm$0.028 \\
& Gemini-S/TReCS & Si-1 & 7.73  & 30.98$\pm$0.16 \\
& Gemini-S/TReCS & Si-3 & 9.69  & 28.82$\pm$0.18 \\
& Gemini-S/TReCS & Si-5 & 11.66 & 25.24$\pm$0.18 \\
& Gemini-S/TReCS & Qa & 18.30   & 19.42$\pm$0.22 \\
& Gemini-S/TReCS & Qb &  24.56  & 10.89$\pm$0.23 \\ \tableline
{\it{Day 1616}} & IRSF/SIRIUS    &   J  & 1.22  & 0.51$\pm$0.05  \\ 
& IRSF/SIRIUS    &   H  & 1.65  & 0.54$\pm$0.06  \\
& IRSF/SIRIUS    &  Ks  & 2.16  & 0.66$\pm$0.12  \\
& Gemini-S/TReCS & Si-1 & 7.73  & 23.29$\pm$0.14 \\
& Gemini-S/TReCS & Si-3 & 9.69  & 26.43$\pm$0.16 \\
& Gemini-S/TReCS & Si-5 & 11.66 & 23.36$\pm$0.16 \\
& Gemini-S/TReCS &  Qa  & 18.30 & 18.32$\pm$0.54 \\
& Gemini-S/TReCS &  Qb  & 24.56 & 12.89$\pm$0.61 \\ \tableline
{\it{Day 1947}} & IRSF/SIRIUS    &   J  & 1.22  &  0.49$\pm$0.05  \\
& IRSF/SIRIUS    &   H  & 1.65  &  0.47$\pm$0.05 \\
& IRSF/SIRIUS    &  Ks  & 2.16  &  0.57$\pm$0.06 \\
& Gemini-S/TReCS & Si-1 & 7.73  & 16.53$\pm$0.04 \\
& Gemini-S/TReCS & Si-3 & 9.69  & 17.88$\pm$0.04 \\
& Gemini-S/TReCS & Si-5 & 11.66 & 16.56$\pm$0.02 \\
& Gemini-S/TReCS &  Qa  & 18.30 & 14.69$\pm$0.06 \\
& Gemini-S/TReCS &  Qb  & 24.56 &  9.59$\pm$0.20 \\
\enddata
\end{deluxetable}

\section{The Near Infrared Spectrum of V1280 Sco at  Day 940\label{sect4}}
\begin{figure}[htbp!]
\begin{center}
\includegraphics{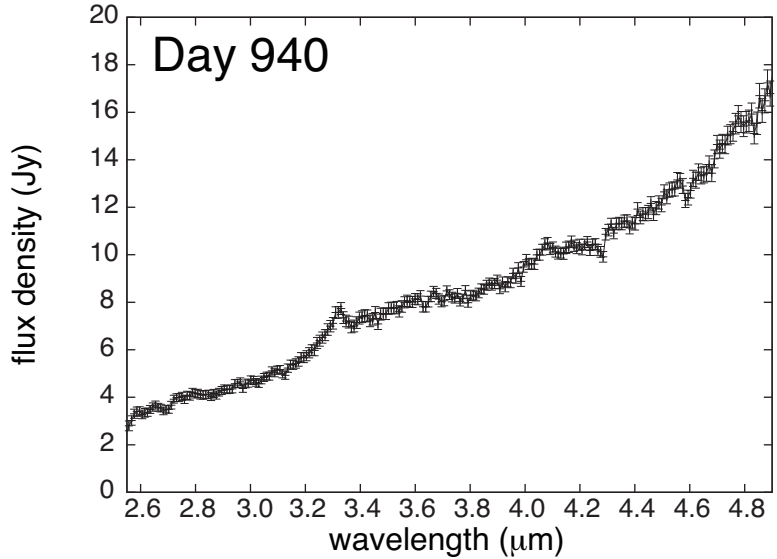}
\caption{The near-infrared spectrum of V1280 Sco on Day 940 obtained with AKARI/IRC observations. \label{V1280SCO_AKARI_NG_D940} }
\end{center}
\end{figure}

Figure~\ref{V1280SCO_AKARI_NG_D940} shows the near-infrared spectrum of V1280 Sco on Day 940 obtained with AKARI/IRC. The spectrum is characterized by strong red continuum emission with a small unidentified infrared (UIR) emission feature at 3.3~$\mu$m. Small dents at 4.26~$\mu$m and 4.6~$\mu$m may be the absorption features due to CO$_2$ gas and CO gas, respectively. The properties of the carriers of the emission and/or absorption features are discussed in \S~\ref{sect6}. 

\section{Temporal Evolution of Infrared SEDs of V1280 Sco \label{sect5}}
\subsection{Temperature and Luminosity Evolution of the White Dwarf}
Adopting $M_{{\rm WD}}=0.6M_{\odot}$ as the mass of the white dwarf in V1280 Sco \citep[e.g.,][]{hou10,nai12}, the temperature and luminosity of the white dwarf on Day 150 after the outburst are estimated to be $T_{\rm WD}=2.5\times 10^{4}$~K and $L_{\rm WD}=1.6\times 10^{4}L_{\odot}$ assuming the chemical composition in the envelope of $X=0.35$, $Y=0.33$, C$+$O$=0.30$, where X is the fractional abundance of hydrogen, Y is that of helium, and C$+$O$=0.30$ is the combined abundance of carbon and oxygen, respectively \citep{kat94}. When the nova evolves to the nebular phase, the temperature and luminosity of the white dwarf reaches $T_{\rm WD}=1.0\times 10^{5}$~K and $L_{\rm WD}=1.8\times 10^{4}L_{\odot}$. In the following analyses, we adopt these latter values for the white dwarf on Days 1272, 1616 and 1947. We note that those luminosity values are significantly larger than that estimated from the core mass to luminosity relation for white dwarfs given by \citet{pac71}. However, the relationship by \citet{pac71} is applicable only to the case where the temperature of the C$+$O core is much higher than in novae and, therefore, is not appropriate for our analysis.

\subsection{Infrared SED of V1280 Sco on Day 150}
\subsubsection{Analysis Based on Optically-thin Dust Model Fitting}
Figure~\ref{SED_NL_D150_FIT} shows the near- to mid-infrared SED of V1280 Sco on Day 150. It is dominated by the thermal dust emission with a small absorption structure at around 9.7~$\mu$m due to silicate dust. Although the origin of the silicate dust is not clear, it is likely to be predominantly associated with interstellar silicate dust located along the line of sight towards V1280 Sco since neither apparent emission features from newly formed silicate nor infrared light echoes from the circumstellar silicate have been observed between Days 23 and 145 \citep{che08}. To our knowledge, dust components that have ever been identified in CO novae are amorphous carbon, hydrogenated amorphous carbon, SiC and silicates \citep{eva12}. Taking into account the relatively smooth spectral shape except for a small absorption structure by interstellar silicate, we expect that amorphous carbon is responsible for the infrared continuum emission observed in V1280 Sco on Day 150.

\subsubsection{Can an Optically Thin Dust Emission Model Explain the Observed SED on Day 150? \label{SED_D150}}
\begin{figure}[htbp!]
\begin{center}
\includegraphics{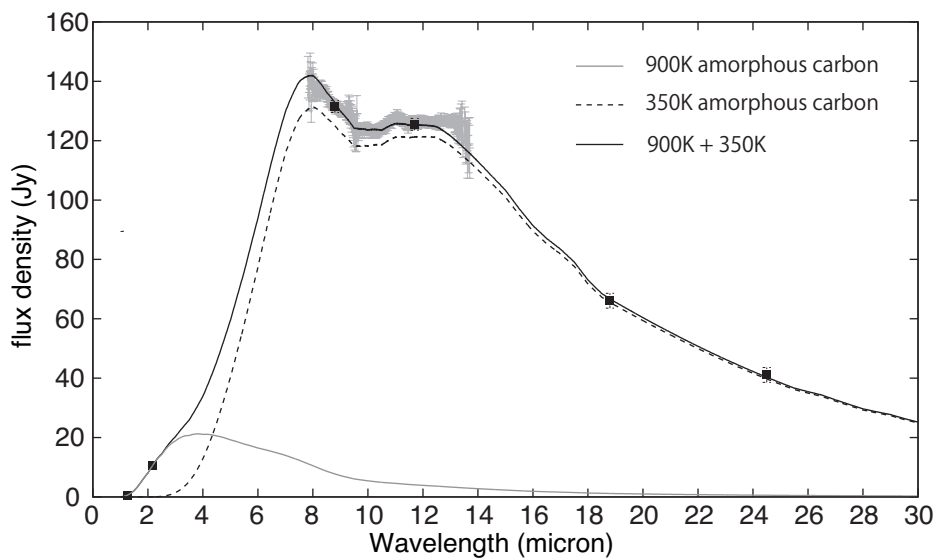}
\caption{Near to mid-infrared SED of V1280 Sco on Day 150 constructed by the photometric data in the near- to mid-infrared ({\it{solid square}}) and the N-band low resolution spectroscopic data ({\it{gray pluses}}) \label{SED_NL_D150}. Assuming optically thin emission modeled with Eq.~(\ref{eq2}), the best-fit spectrum to reproduce the observed near- to mid-infrared SED on Day 150 is shown with thick black line. The optically thin emission from the amorphous carbon components of 900~K and 350~K are shown with gray line and dashed line, respectively, taking account of the foreground silicate extinction of $\tau_{9.7}=0.24$. 
\label{SED_NL_D150_FIT} }
\end{center}
\end{figure}

Assuming that spherical amorphous carbon dust grains with a uniform radius $a$ and a total mass $M_i$ are located at a distance $R$ from the observer, and that they give off optically thin thermal emission with an equilibrium temperature $T_i$, the observed flux density is 
\begin{eqnarray}
f^i_{\nu}(\lambda)=M_i \left( \frac{4}{3}\pi \rho_{a.car.} a^3 \right)^{-1} \pi B_{\nu}(\lambda, T_i) Q^{abs}_{a.car.}(\lambda) \left(\frac{a}{D}\right)^2, \label{eq1}
\end{eqnarray}
where $\rho_{a.car.}$ is the density of an amorphous carbon dust particle (we use $\rho_{a.car.}=1.87$g cm$^{-3}$) and $Q^{abs}_{a.car.}(\lambda)$ is the absorption efficiency of amorphous carbon of a radius $a$ \citep[ACAR sample;][]{zub96}. The optically-thin emission from multi-temperature amorphous carbon components with foreground silicate extinction is given by 
\begin{eqnarray}
f^{model}_{\nu}(\lambda)=\sum^{N}_{i=1} f^i_{\nu} (\lambda) \exp\left\{-\tau_{9.7}\left(\frac{Q^{abs}_{a.sil.}(\lambda)}{Q^{abs}_{a.sil.}(9.7 \mu m)} \right)\right\}, \label{eq2}
\end{eqnarray}
where $N$ is the number of amorphous carbon components with different representative temperatures, $Q^{abs}_{a.sil.}(\lambda)$ is the absorption efficiency of astronomical silicate of a radius $a$ \citep{dra85}, and $\tau_{9.7}$ is the optical depth of foreground silicate dust at 9.7~$\mu$m. The observed near- to mid-infrared SED on Day 150 is reproduced well by the optically thin emission from amorphous carbon components at two different temperatures (i.e., $N=2$) 350~K and 900~K with a foreground silicate extinction with $\tau_{9.7}=0.24\pm0.01$ (see Fig.~\ref{SED_NL_D150_FIT}). The best-fit parameters are summarized in Table~\ref{tbl_param_thin_d150}. 

$A_V=1.2$ mag is inferred from \citet{mar06} towards the direction of V1280 Sco. Assuming the ratio of $A_V/\tau_{9.7}=18.5\pm$1.5 \citep{roc84}, the interstellar extinction expected at 9.7~$\mu$m toward the direction of V1280 Sco is $\tau_{9.7}=0.065$, which is smaller than the value $\tau_{9.7}=0.24$ obtained from our SED analysis. This discrepancy may be attributed to the uncommon composition of dust grains intervening between the V1280 Sco and us, given that the value of $A_V/\tau_{9.7}$ is sensitive to the abundance ratio of silicate dust to carbon dust. Because no apparent light echoes containing silicate emission have been reported between Days 23 and 145 by \citet{che08}, it is less likely that the silicate that contribute in producing absorption structure at 9.7µm on Day 150 is associated with V1280 Sco. In our SED analysis on Days 1272, 1616 and 1947 in \S~\ref{SED_RT}, therefore, we assume that the extinction of $\tau_{9.7}=0.24$ is fully associated with interstellar silicate dust located in the line of sight towards V1280 Sco. We note that the results of the SED analysis in \S~\ref{SED_RT} are not significantly altered whatever value between $\tau_{9.7}=0.24$ and $\tau_{9.7}=0.065$ we adopt as the foreground silicate extinction in the direction towards V1280 Sco.

\begin{table}[htbp!]
\begin{center}
\caption{
The dust properties in V1280 Sco on Day 150 assuming an optically thin emission model. \label{tbl_param_thin_d150}
}
\begin{tabular}{ccc}
\tableline\tableline
                  & Component 1 ($i=1$)      & Component 2 ($i=2$)\\
\tableline
Composition       & Amorphous Carbon & Amorphous Carbon \\
Temperature $T_i$ & $900\pm10$~K      & $350\pm10$~K \\
Mass $M_i$        & $3.2\times 10^{-9}$M$_{\odot}$ & $1.0\times 10^{-6}$M$_{\odot}$ \\ 
\tableline
\end{tabular}
\end{center}
\end{table}

The blackbody angular radius $\theta_{bb}$ in arcsecond for the $T_{i=2}=350$~K component is given by
\begin{eqnarray}
\theta_{bb}=2.06\times 10^5  \left( \int^{\infty}_{0} f^{i=2}_{\nu}(\lambda) d\nu \right)^{1/2} \left( \sigma T_{i=2}^4 \right)^{-1/2}  \label{eq3}
\end{eqnarray}
where $f^{i=2}_{\nu}(\lambda)$ is the observed flux density carried by 350~K component and $\sigma$ is the Stefan-Boltzmann constant. We get $\theta_{bb}=0".054$, which is just slightly below the radius of the emitting region of 350~K component (i.e. 0".065) obtained from the deconvolved 24.5~$\mu$m image on Day 150. Therefore, the 350~K dust shell must be both optically thick and spatially unresolved. In this case, the dust mass of amorphous carbon becomes $2.7\times 10^{-6}$M$_{\odot}$ by assuming that the Plank mean-emission cross section for a carbon grain of radius $a$ is proportional to $aT_{i=2}^2$ \citep{gil74,geh80} and that the density of amorphous carbon is $\rho_{a.car.}=1.87$g cm$^{-3}$.

\subsection{Infrared SEDs of V1280 Sco on Days 1272, 1616 and 1947 \label{SED_RT}}
\subsubsection{Analysis Based on Optically-thin Dust Model Fitting}
\begin{figure}
\begin{center}
\includegraphics[width=0.5\linewidth]{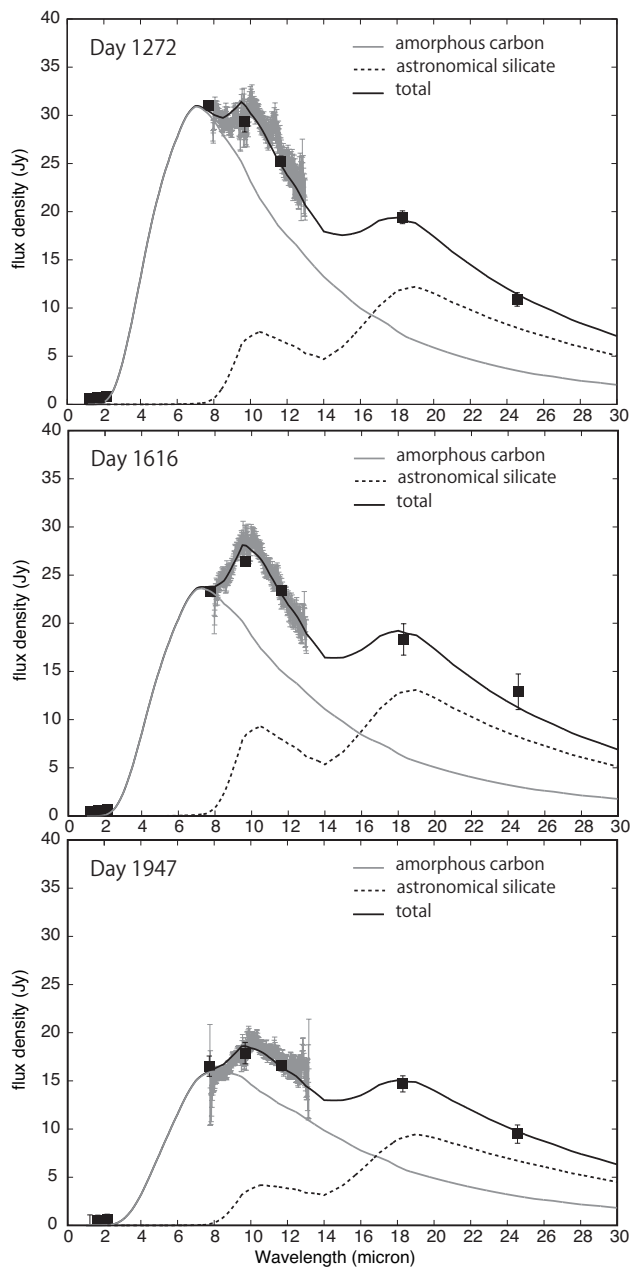}
\caption{Near to mid-infrared SED of V1280 Sco at 1272d(top), 1616d(middle) and 1947d(bottom). The photometric values in the near- to mid-infrared are shown with black squares and the N-band low-resolution spectrum is shown with gray pluses. The best fit model curve assuming optically thin emission from amorphous carbon and astronomical silicate given by eq.~\ref{eq6} is shown with solid black line in each panel. The gray line and broken line in each panel indicate the spectra of emission carried by amorphous carbon and astronomical silicate, respectively. \label{SED_FIT_THIN}
  }
\end{center}
\end{figure}

A notable point in the near- to mid-infrared SEDs of V1280 Sco on Days 1272, 1616 and 1947 (see Fig~\ref{SED_FIT_THIN}) is a conspicuous appearance of the 9.7~$\mu$m and 18~$\mu$m amorphous silicate features in emission, which were not present in the SED obtained on Day 150 (see Fig~\ref{SED_NL_D150_FIT}).

To investigate the temporal evolution of properties of constituent dust, a simple SED fitting was carried out at each epoch based on a two-component (astronomical silicate and amorphous carbon) model, a single equilibrium temperature is assumed for each dust component. Assuming optically thin geometry in the infrared, the observed flux density is given by
\begin{eqnarray}
f^{model}_{\nu}(\lambda)&=&M_{a.car.} \left( \frac{4}{3}\pi \rho_{a.car.} a^3 \right)^{-1} \pi B_{\nu}(\lambda, T_{a.car.}) Q^{abs}_{a.car.}(\lambda) \left(\frac{a}{D}\right)^2 \\ \nonumber
&& + M_{a.sil.} \left( \frac{4}{3}\pi \rho_{a.sil.} a^3 \right)^{-1} \pi B_{\nu}(\lambda, T_{a.sil.}) Q^{abs}_{a.sil.}(\lambda) \left(\frac{a}{D}\right)^2 \label{eq6},
\end{eqnarray}
where $T_{a.car.}$ and $T_{a.sil.}$ are the equilibrium temperature, $M_{a.car.}$ and $M_{a.sil.}$ are the dust mass, $\rho_{a.car.}$ (1.87 g cm$^{-3}$) and $\rho_{a.sil.}$ (3.3 g cm$^{-3}$) are the density of an amorphous carbon ($a.car.$) and astronomical silicate ($a.sil.$) dust particle, respectively. We note that our SED analysis requires the optical depth of ISM silicates at 9.7~$\mu$m to be $\tau_{9.7}=0.24\pm 0.01$ to account for the small absorption structure in the spectrum of V1280 Sco on Day 150. Assuming that this extinction is fully of interstellar origin, the best fit parameters obtained for Days 1272, 1616 and 1947 based on the two-component model are summarized in Table~\ref{tbl_parameters_M_T}. Here we assume that the dust size is smaller than 1~$\mu$m and thus $Q^{abs}_{a.car.}(\lambda)$ and $Q^{abs}_{a.sil.}(\lambda)$ are approximately proportional to $a$ in the mid-infrared wavelength range. The obtained dust masses $M_{a.car.}$ and $M_{a.sil.}$ are, therefore, almost independent of $a$.

\begin{table}[htbp!]
\begin{center}
\caption{
The dust properties in V1280 Sco on Days 1272, 1616 and 1947. \label{tbl_parameters_M_T}
}
\scriptsize
\begin{tabular}{lccccccc}
\tableline \tableline
  & \multicolumn{3}{c}{Amorphous Carbon} & & \multicolumn{3}{c}{Astronomical Silicate} \\
\cline{2-4}
\cline{6-8}
 Epoch (days)        &     1272      &    1616    &    1947  &   &   1272        & 1616          &  1947     \\ \tableline
Temperature (K)     &   540$\pm$10    & 500$\pm$10  & 440$\pm$10 &  &   220$\pm$10   & 230$\pm$10     &  210$\pm$10     \\
Mass (10$^{-7}$$M_{\odot}$)   &   0.79$^{+0.05}_{-0.05}$  &   0.79$^{+0.05}_{-0.04}$   &   0.89$^{+0.12}_{-0.10}$ &  &   8.4$^{+2.8}_{-1.0}$   &  7.1$^{+3.2}_{-1.2}$    & 7.5$^{+3.8}_{-2.0}$     \\
\tableline
\end{tabular}
\end{center}
\end{table}

\begin{figure}
\begin{center}
\includegraphics{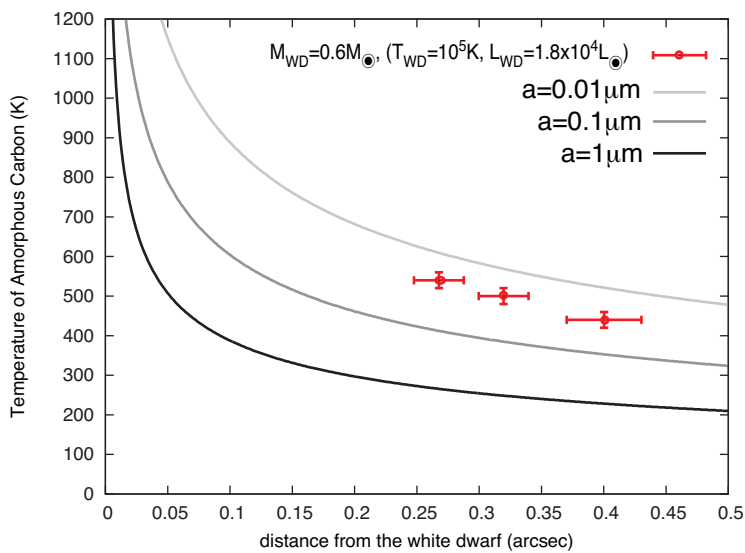}
\caption{Plots of the temperature of amorphous carbon with different radii of $a=0.01$~$\mu$m, $0.1$~$\mu$m, and $1.0$~$\mu$m calculated as a function of the distance $r$ from the white dwarf of $M_{\rm WD}=0.6M_{\odot}$ modeled with $(T_{\rm WD}, L_{\rm WD})=(1.0\times 10^{5}, 1.8\times 10^4L_{\odot})$. The datapoints obtained as a result of the SED fitting at each epoch on Days 1272, 1616, 1947 are shown with red crosses, where the semi-major axis radius of the deconvolved emitting region modeled with the elliptic Gaussian for the Si-1 is adopted to indicate the location of amorphous carbon component.  \label{T_EV_AC}}
\end{center}
\end{figure}

The temporal evolution of amorphous carbon properties is characterized by a decrement in the temperature without significant mass evolution at those three epochs. This suggests the kinematic behavior of the amorphous carbon dust, which travels farther from the white dwarf without being destroyed. The projected distance corresponding to the semi-major axis of the deconvolved emitting region modeled with an elliptic Gaussian in the Si-1 band, which is supposed to trace amorphous carbon component in our photometric dataset (Fig.~\ref{SED_FIT_THIN}), is 0".27$\pm$0".02 on Day 1272, 0".32$\pm$0".02 on Day 1616, and 0".40$\pm$0".04 on Day 1947. If the spread of dust is in the shape of a bipolar nebulae and is viewed edge on \citep[e.g.,][]{che12}, the size of the emitting region corresponds to the average distance to which the dust had traveled from the white dwarf. Figure~\ref{T_EV_AC} shows temperatures of amorphous carbon grains of different radii $a=0.01$~$\mu$m, $0.1$~$\mu$m, and $1.0$~$\mu$m as a function of distance from the white dwarf. We base our calculations on the same WD properties as before, i.e., $M_{\rm WD}=0.6M_{\odot}$, $T_{\rm WD}=1.0\times 10^{5}$~K, and $L_{\rm WD}=1.8\times 10^4 $~$L_{\odot}$, and further assume that these values all remain constant on Days 1272, 1616, and 1947. Three datapoints obtained by fitting the SED are located between the model curves for $a=0.01\mu$m and 0.1$\mu$m. Therefore, with our two-component model that assumes a single equilibrium temperature for each, the grain size of amorphous carbon dust formed in the nova ejecta is estimated to be in a range from $a=0.01$~$\mu$m to 0.1~$\mu$m. We note, however, that a temperature distribution in each dust component is expected but not taken into consideration in this simple model. A more accurate estimate of the grain size of amorphous carbon assuming an appropriate dust geometry is discussed in \S~\ref{SED_RT}.

The Qa and Qb images taken on Day 1947 (see Fig.~\ref{IMG1947}) are dominated by the astronomical silicate emission, as we have shown above (see Fig.~\ref{SED_FIT_THIN}), and they exhibit an elongated shape along PA$\sim 20$deg. The similarity between the image shapes taken in Qa and Qb bands and those in Si-1, S-3 and Si-5 bands may indicate that both the carbon and silicate dust components reside predominantly in a dusty bipolar nebulae \citep{che12}. The projected distance corresponding to the semi-major axis of the deconvolved emitting region modeled with an elliptic Gaussian is 0".40$\pm$0".03 in the Qa band and 0".45$\pm$0".04 in the Qb band on Day 1947. 

\begin{figure}
\begin{center}
\includegraphics{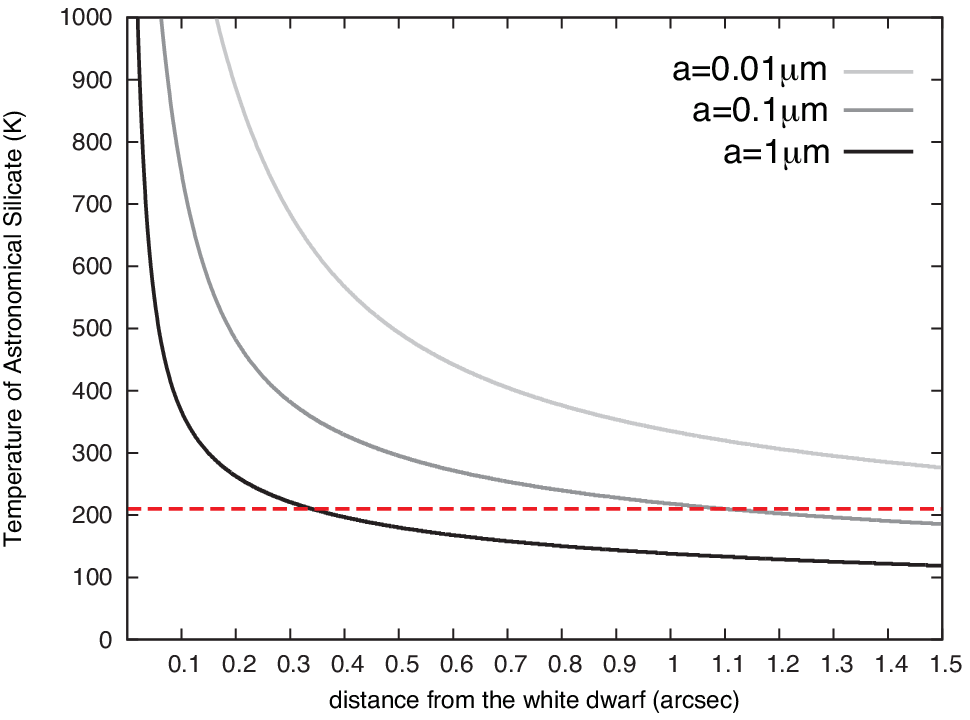}
\caption{Plots of the temperature of astronomical silicate with different radii of $a=0.01$~$\mu$m, $0.1$~$\mu$m, and $1.0$~$\mu$m calculated as a function of the distance $r$ from the white dwarf of $M_{\rm WD}=0.6M_{\odot}$ modeled with $(T_{\rm WD}, L_{\rm WD})=(1.0\times 10^{5}, 1.8\times 10^4$~$L_{\odot})$. The broken red line indicates the temperature of the astronomical silicate component obtained as a result of the SED fitting on Day 1947 assuming optically thin dust emission model.
 \label{T_EV_AS}}
\end{center}
\end{figure}

Figure~\ref{T_EV_AS} shows equilibrium temperatures of optically thin astronomical silicate as a function of distance from the white dwarf calculated for different radii $a=0.01$~$\mu$m, $0.1$~$\mu$m, and $1.0$~$\mu$m. Astronomical silicate dust achieves an equilibrium dust temperature of 210~K at a distance of 1".1 from the white dwarf for $a=0.1$~$\mu$m and 0".34 for $a=1.0$~$\mu$m. Comparing these values with the semi-major axes of the mid-infrared emitting region in the Qa and Qb bands derived above, we conclude that the typical grain size of astronomical silicate around V1280Sco must be in a range from 0.1 to 1.0~$\mu$m. In the next section, the results of the SED analyses taking account of the dust geometry are discussed to examine more detailed physical properties of circumstellar dust around V1280 Sco. Then, the origin of each dust species is discussed by investigating the temporal evolution of the resultant geometric and physical parameters of circumstellar dust around V1280 Sco among the three epoch in \S~\ref{DUST_HISTORY}.
\pagebreak

\subsubsection{Dust SED Analysis  \label{SED_RT}}
\begin{figure}
\begin{center}
\includegraphics[width=0.8\linewidth]{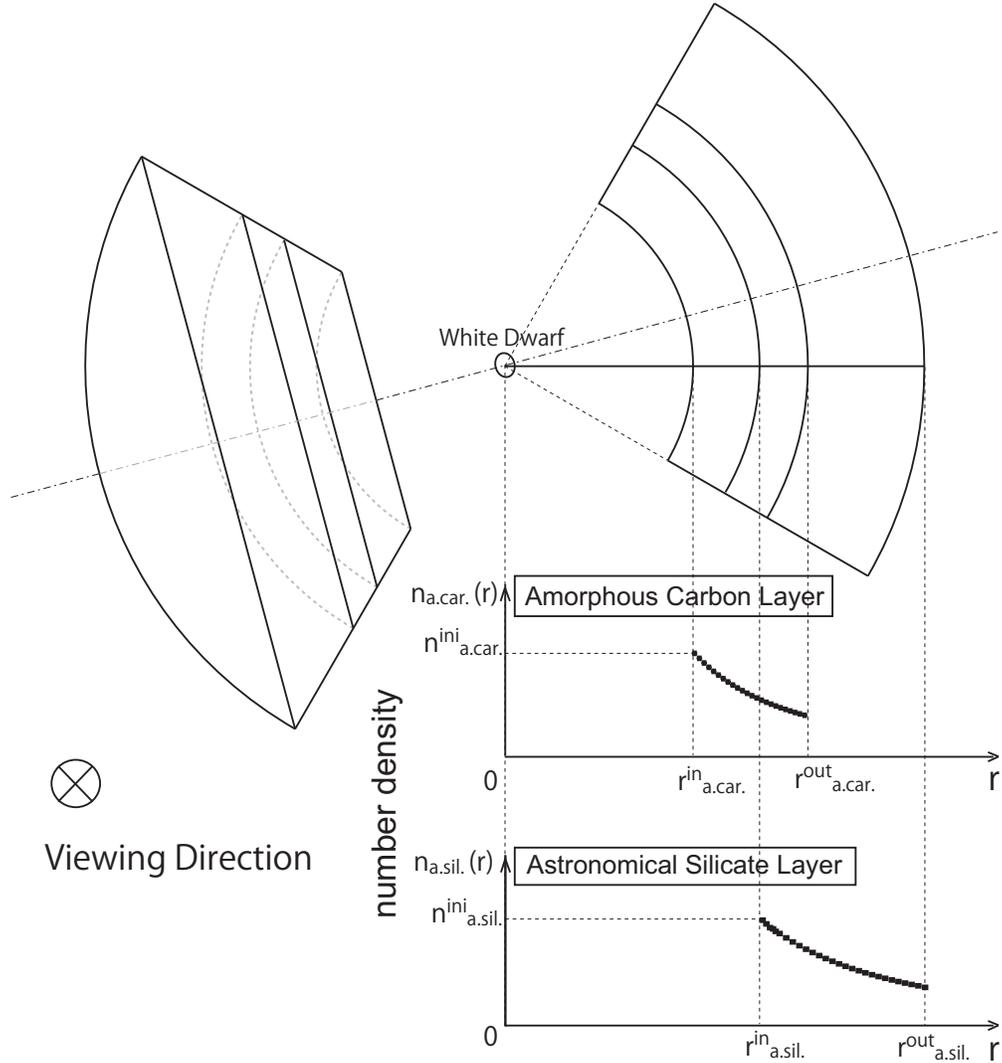}
\caption{The schematic view of dust geometry assumed in V1280 Sco on Days 1272, 1616, and 1947. Amorphous carbon and astronomical silicate are assumed to be distributed in bipolar cones with a common opening angle $\phi$. Each component is confined within the volume enclosed by the inner and outer walls, where the density distribution is assumed to be proportional to $r^{-2}$. The viewing direction is from the surface to the back of the paper.
\label{geometry}
}
\end{center}
\end{figure}

The elongated shape of V1280 Sco along PA$\sim20$deg seen both in the N- and Q-band images on Day 1947 suggests that the mid-infrared emission is dominated by carbon and silicate dust grains in a bipolar nebula. This leads us to assume a simple geometry, in which both of the amorphous carbon and astronomical silicate components are distributed in bipolar cones with a common opening angle $\phi$ (see Fig.~\ref{geometry}) and is viewed edge on. 

This geometry is consistent with the description of the structure of the ejected shells given by \citet{nai13}. We assume the density distributions of amorphous carbon [$n_{a.car.}(r)$] and astronomical silicate [$n_{a.sil.}(r)$] to be 
\begin{eqnarray}
n_{a.car.}(r)=\left\{ \begin{array}{ll}
     0 & (r<r^{in}_{a.car.}) \\
     n^{ini}_{a.car.} \times \left( \frac{\displaystyle r}{\displaystyle r^{in}_{a.car.}} \right)^2 & (r^{in}_{a.car.}< r < r^{out}_{a.car.}) \\
     0 & (r^{out}_{a.car.}<r) 
\end{array}
\right.  \label{eq_acar_density}
\end{eqnarray}
and
\begin{eqnarray}
n_{a.sil.}(r)=\left\{ \begin{array}{ll}
     0 & (r<r^{in}_{a.sil.}) \\
     n^{ini}_{a.sil.} \times \left( \frac{\displaystyle r}{\displaystyle r^{in}_{a.sil.}} \right)^2 & (r^{in}_{a.sil.}< r < r^{out}_{a.sil.}) , \\
     0 & (r^{out}_{a.sil.}<r) 
\end{array}
\right.  \label{eq_asil_density}
\end{eqnarray}
respectively, as a function of distance $r$ from the white dwarf, where $r^{in}_{a.car.}$ and $r^{in}_{a.sil.}$ are the inner wall radii, $r^{out}_{a.car.}$ and $r^{out}_{a.sil.}$ are the outer wall radii, and $n^{ini}_{a.car.}$ and $n^{ini}_{a.sil.}$ are the initial number densities at the position of the inner wall for amorphous carbon and astronomical silicate, respectively. Analyses are made independently for two different spherical grain radii $a_{a.car.}=a_{a.sil.}=0.1~\mu$m and $a_{a.car.}=a_{a.sil.}=0.01~\mu$m. 

For such a geometric configuration and density distributions, the flux density from the white dwarf, $f^{\rm WD}_{\nu}(\lambda,\zeta)$, at $r=\zeta$ is given by
\begin{eqnarray}
f^{{\rm WD}}_{\nu}(\lambda,\zeta)&\simeq& \left(\frac{L_{\rm WD}}{\sigma T_{{\rm WD}}^4} \right) \pi B_{\nu}(T_{{\rm WD}}) \frac{1}{4\pi \zeta^2} \\ \nonumber  
&& \exp\left\{ -\pi a_{a.car.}^2 Q^{abs}_{a.car.}(\lambda)\int^{\zeta}_{0} n_{a.car.}(r)dr\right\} \\ \nonumber 
&& \exp\left\{ -\pi a_{a.sil.}^2 Q^{abs}_{a.sil.}(\lambda)\int^{\zeta}_{0} n_{a.sil.}(r)dr\right\}.
\end{eqnarray}

We do not take into account backward scattering for simplification and assume the same stellar parameters as before for the white dwarf on Days 1272, 1616, and 1947 (i.e., $T_{\rm WD}=1.0\times 10^{5}$~K and $L_{\rm WD}= 1.8\times 10^4L_{\odot}$). The temperature of a dust grain, $T_{\chi}(\zeta)$, where $\chi=a.car.$ or $a.sil.$ in our case, in local radiative equilibrium at $r=\zeta$ satisfies 
\begin{eqnarray}
\int^{\infty}_{0} 4\pi a_{\chi}^2 Q^{abs}_{\chi}(\lambda) \pi B_{\nu}(T_{\chi}(\zeta)) d\nu = \int^{\infty}_{0} \pi a_{\chi}^2 Q^{abs}_{\chi}(\lambda) f^{{\rm WD}}_{\nu}(\lambda,\zeta) d\nu.
\end{eqnarray}
The absorption coefficients of astronomical silicate of radius $a$ at shorter wavelengths ($\lambda<0.1$~$\mu$m) are calculated from \citet{dwe96} and those at longer wavelengths ($\lambda>0.1$~$\mu$m) are quoted from \citet{dra85}. The heating due to the energy absorption of thermal infrared emission from other dust particles is not taken into account.

Assuming that V1280 Sco is viewed edge, an infrared spectrum produced by amorphous carbon and astronomical silicate in bipolar lobes of V1280 Sco is calculated for a geometric configuration characterized by 7 free parameters; $n^{ini}_{a.car.}$, $n^{ini}_{a.sil.}$, $r^{in}_{a.car.}$, $r^{in}_{a.sil.}$, $r^{out}_{a.car.}$, $r^{out}_{a.sil.}$, and $\phi$. The optical depth at 9.7$\mu$m, $\tau_{9.7}=0.24\pm 0.01$, due to interstellar silicate towards V1280 Sco is fixed. Least-squares fitting is carried out to obtain a set of 7 best-fit parameters on Days 1272, 1616 and 1947. Datapoints used for the model fitting were selected to exclude any possible dust emission features and emission lines, in the N-band low-resolution spectra in addition to the N- and Q-band photometric data.

\begin{figure}
\begin{center}
\includegraphics[width=0.8\linewidth]{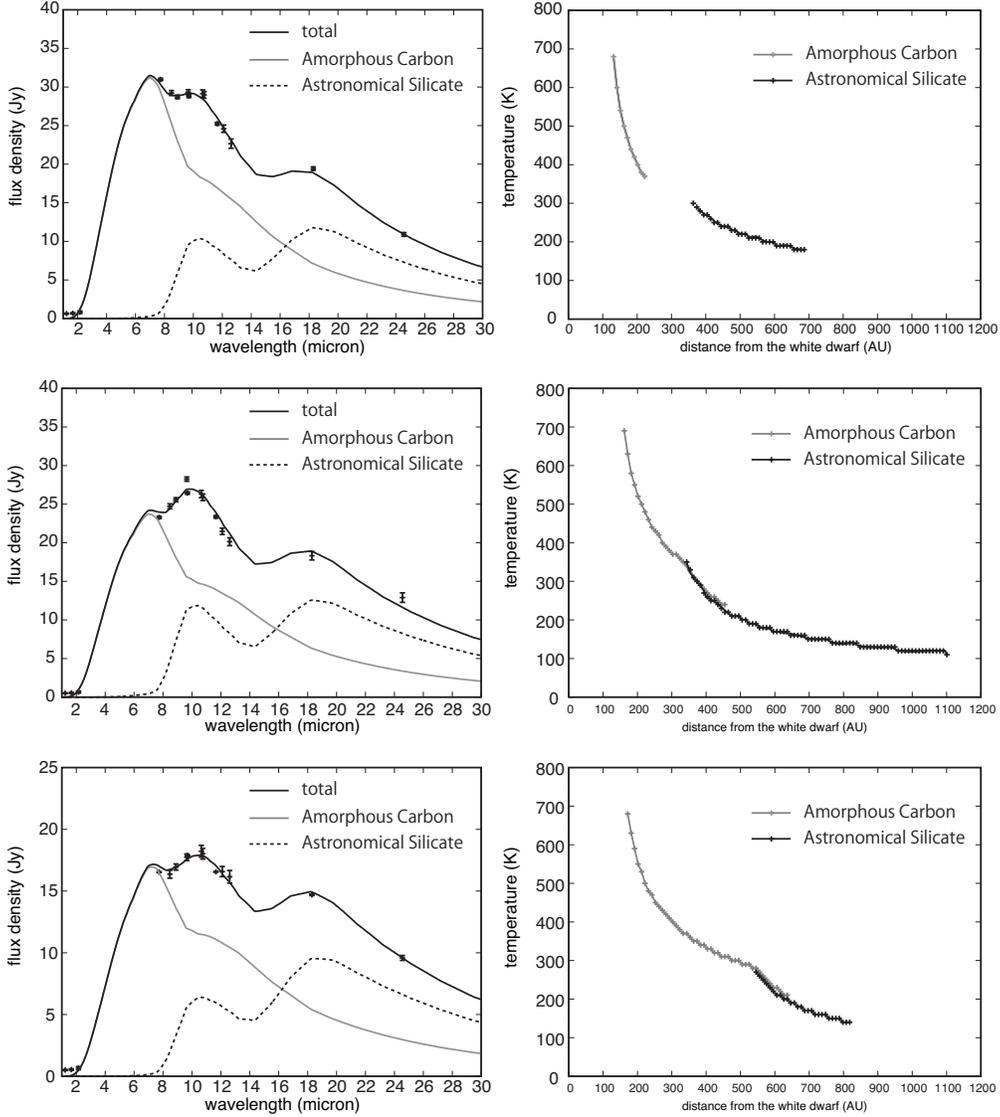}
\caption{The datapoints used for the model fitting were constructed from the continuum points, which were selected to exclude any possible dust emission features and emission lines, in the N-band low-resolution spectra in addition to the N-band and Q-band photometric data. The best fit curves obtained as a result of the SED fitting at each epoch of Days 1272, 1616, and 1947 assuming a single grain size of $a_{a.car.}=a_{a.sil.}=0.01~\mu$m are shown in the right panels with solid black line. The gray line and broken line in each left panel indicate the spectra of emission carried by amorphous carbon and astronomical silicate, respectively. Plots of the temperature of amorphous carbon and astronomical silicate as a function of the distance from the white dwarf obtained as a result of the SED fitting at each epoch on Days 1272, 1616 and 1947 are shown in the right panels.
 \label{fig_best_fit_SED_RT_1}
  }
\end{center}
\end{figure}

\begin{figure}
\begin{center}
\includegraphics[width=0.8\linewidth]{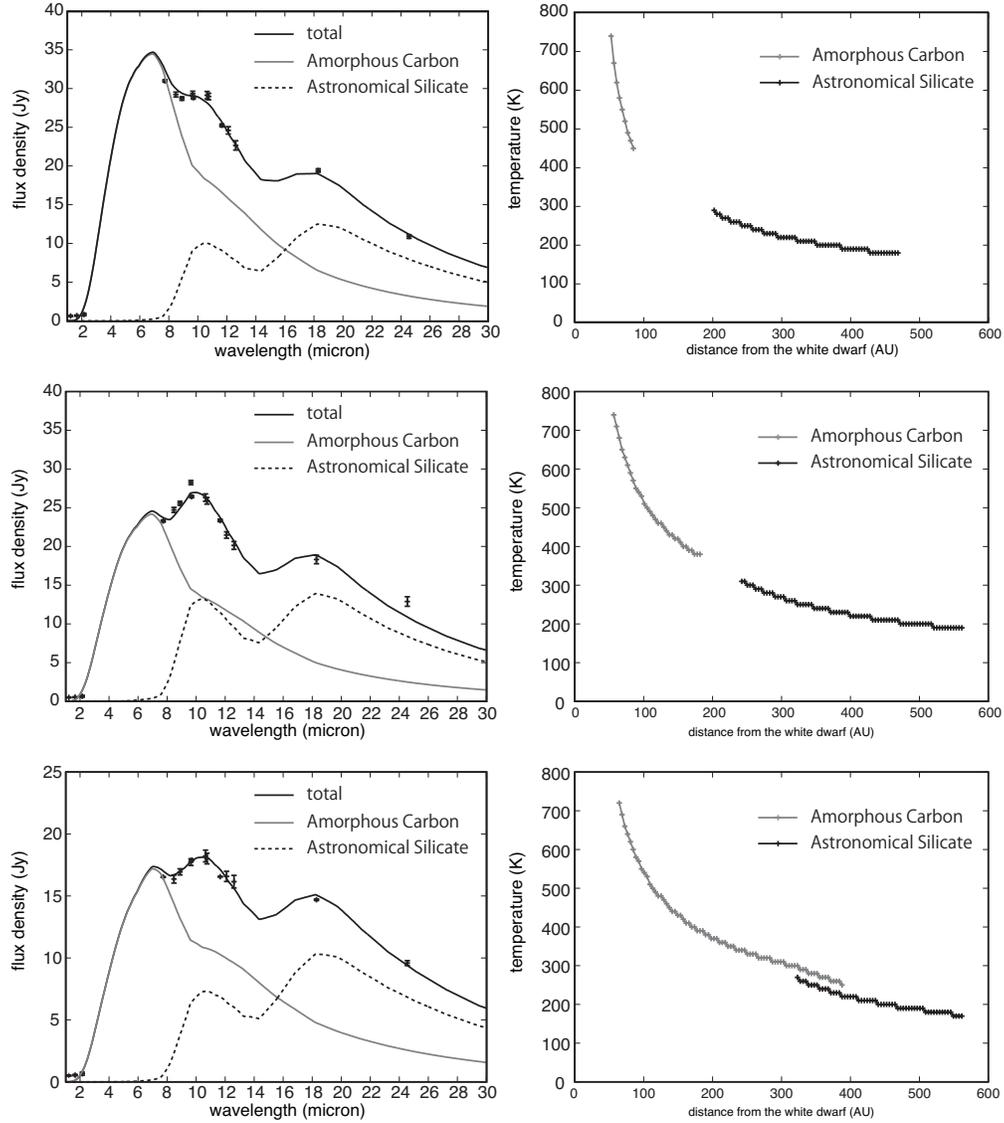}
\caption{The same as Fig.~\ref{fig_best_fit_SED_RT_1} but for $a_{a.car.}=a_{a.sil.}=0.1\mu$m.
 \label{fig_best_fit_SED_RT_2}
  }
\end{center}
\end{figure}

The best-fit parameters obtained on Days 1272, 1616 and 1947 calculated for the two cases $a_{a.car.}=a_{a.sil.}=0.01~\mu$m and $a_{a.car.}=a_{a.sil.}=0.1~\mu$m are summarized in Table~\ref{tbl_parameters_geometry_1} and Table~\ref{tbl_parameters_geometry_2}, respectively. The model spectra calculated with those best-fit parameters are shown in the left panels of Figs.~\ref{fig_best_fit_SED_RT_1} and \ref{fig_best_fit_SED_RT_2}. Temperatures of amorphous carbon and astronomical silicate are shown as a function of distance from the white dwarf in the right panels of Figs~\ref{fig_best_fit_SED_RT_1} and \ref{fig_best_fit_SED_RT_2}. In both cases, the temporal evolution of infrared SED of V1280 Sco at these three epochs can be explained by a simple scenario in which amorphous carbon and astronomical silicate dust grains are moving away from the white dwarf. 

\begin{table}[htbp!]
\begin{center}
\caption{
The best-fit parameters obtained for $a_{a.car.}=a_{a.sil.}=0.01~\mu$m \label{tbl_parameters_geometry_1}
}
\scriptsize
\begin{tabular}{lccc}
\tableline \tableline
Epoch (days)            
&      1272        
&      1616         
&      1947        \\ \tableline
opening angle; $\phi$ [deg] 
&  24.0$\pm$0.5    
&  18.0$\pm$0.5    
&  15.0$\pm$0.5   \\
Amor. Car. initial number density; $n^{ini}_{a.car.}$ [cm$^{-3}$] 
& 2.5$^{+0.3}_{-0.3}$$\times$$10^{-3}$ 
& 1.3$^{+0.2}_{-0.2}$$\times$$10^{-3}$ 
& 1.0$^{+0.2}_{-0.1}$$\times$$10^{-3}$ \\
Amor. Car. inner radius; $r^{in}_{a.car.}$ [AU] 
& 1.3$^{+0.1}_{-0.1}$$\times$$10^{2}$ 
& 1.5$^{+0.1}_{-0.1}$$\times$$10^{2}$
& 1.7$^{+0.1}_{-0.2}$$\times$$10^{2}$ \\
Amor. Car. outer radius; $r^{out}_{a.car.}$ [AU] 
& 2.4$^{+0.3}_{-0.2}$$\times$$10^{2}$ 
& 4.6$^{+0.8}_{-0.6}$$\times$$10^{2}$ 
& 6.4$^{+2.2}_{-0.9}$$\times$$10^{2}$ \\
Astro. Sil. initial number density; $n^{ini}_{a.sil.}$  [cm$^{-3}$] 
& 3.2$^{+0.4}_{-0.3}$$\times$$10^{-4}$ 
& 6.0$^{+0.7}_{-0.6}$$\times$$10^{-4}$ 
& 6.3$^{+0.8}_{-0.6}$$\times$$10^{-4}$ \\
Astro. Sil. inner radius; $r^{in}_{a.sil.}$  [AU]
& 3.6$^{+0.4}_{-0.3}$$\times$$10^{2}$
& 3.5$^{+0.4}_{-0.2}$$\times$$10^{2}$
& 5.5$^{+0.6}_{-0.7}$$\times$$10^{2}$\\
Astro. Sil. outer radius; $r^{out}_{a.sil.}$ [AU]
& 7.0$^{+0.6}_{-0.5}$$\times$$10^{2}$
& 11.1$^{+6.5}_{-2.0}$$\times$$10^{2}$
& 8.3$^{+1.6}_{-0.8}$$\times$$10^{2}$\\
\tableline
\end{tabular}
\end{center}
\end{table}

\begin{table}[htbp!]
\begin{center}
\caption{
The best-fit parameters obtained for $a_{a.car.}=a_{a.sil.}=0.1~\mu$m  \label{tbl_parameters_geometry_2}
}
\scriptsize
\begin{tabular}{lccc}
\tableline \tableline
Epoch (days)            
&      1272        
&      1616         
&      1947        \\ \tableline
opening angle; $\phi$ [deg] 
& 26.0$\pm$0.5    
& 19.5$\pm$0.5    
& 15.0$\pm$0.5   \\
Amor. Car. initial number density; $n^{ini}_{a.car.}$ [cm$^{-3}$] 
& 2.5$^{+0.3}_{-0.3}$$\times$$10^{-5}$ 
& 0.9$^{+0.1}_{-0.1}$$\times$$10^{-5}$ 
& 0.8$^{+0.1}_{-0.1}$$\times$$10^{-5}$ \\
Amor. Car. inner radius; $r^{in}_{a.car.}$ [AU] 
& 5.1$^{+0.2}_{-0.6}$$\times$$10^{1}$ 
& 5.4$^{+0.2}_{-0.4}$$\times$$10^{1}$
& 6.1$^{+0.6}_{-0.2}$$\times$$10^{1}$ \\
Amor. Car. outer radius; $r^{out}_{a.car.}$ [AU] 
& 0.9$^{+0.1}_{-0.1}$$\times$$10^{2}$ 
& 1.8$^{+0.4}_{-0.2}$$\times$$10^{2}$ 
& 3.9$^{+1.2}_{-0.8}$$\times$$10^{2}$ \\

Astro. Sil. initial number density; $n^{ini}_{a.sil.}$  [cm$^{-3}$] 
& 1.3$^{+0.2}_{-0.2}$$\times$$10^{-6}$ 
& 1.1$^{+0.2}_{-0.2}$$\times$$10^{-6}$ 
& 1.6$^{+0.2}_{-0.2}$$\times$$10^{-6}$ \\
Astro. Sil. inner radius; $r^{in}_{a.sil.}$ [AU]
& 2.0$^{+0.3}_{-0.2}$$\times$$10^{2}$
& 2.4$^{+0.2}_{-0.2}$$\times$$10^{2}$
& 3.2$^{+0.2}_{-0.3}$$\times$$10^{2}$\\
Astro. Sil. outer radius; $r^{out}_{a.sil.}$ [AU]
& 4.8$^{+0.5}_{-0.4}$$\times$$10^{2}$
& 5.5$^{+0.5}_{-0.3}$$\times$$10^{2}$
& 5.6$^{+0.4}_{-0.2}$$\times$$10^{2}$\\
\tableline
\end{tabular}
\end{center}
\end{table}


We find that $r^{in}_{a.car.}$ on Days 1272, 1616, and 1947, respectively, is 1.3$\times$10$^2$, 1.5$\times$10$^2$ and 1.7$\times$10$^2$AU when $a_{a.car.}=0.01~\mu$m, and 5.1$\times$10$^1$, 5.4$\times$10$^1$ and 6.1$\times$10$^1$AU when $a_{a.car.}=0.1~\mu$m. The dust expansion velocity of 0.35$\pm$0.03 mas day$^{-1}$ reported by \citet{che08} corresponds to 0.19 AU day$^{-1}$. If the expansion velocity stays constant, the dust will reach at the distances of 2.4$\times$10$^2$, 3.1$\times$10$^2$ and 3.7$\times$10$^2$~AU on Days 1272, 1616 and 1947, respectively. Moreover, the deconvolved images of V1280 Sco taken in the Si-1 band on Days 1272, 1616 and 1947 exhibit elongated structures along PA$\sim20$deg and are modeled with elliptical Gaussian profiles with the semi-major axis lengths of 0".27, 0".32 and 0".4, which correspond to 1.5$\times$10$^2$, 1.8$\times$10$^2$ and 2.2$\times$10$^2$~AU, respectively. Judging by the location of amorphous carbon dust in our model, the representative size of amorphous carbon dust produced around V1280 Sco is more likely to be $0.01~\mu$m rather than $0.1~\mu$m. We note that the uncertainty in the adopted distance towards the V1280 Sco can directly affect the distance between the dust and the white dwarf constrained from the observations. However, even if we adopt the shortest distance of $D=0.63$~kpc towards V1280 Sco estimated by \citet{hou10}, the semi-major axis lengths of the deconvolved image observed on Days 1272, 1616, and 1947 correspond to 0.9$\times$10$^2$, 1.0$\times$10$^2$ and 1.3$\times$10$^2$~AU, respectively, which are still much larger than the inner wall radii of amorphous carbons obtained for $a_{a.car.}=0.1~\mu$m and, thus, the representative size of amorphous carbon dust is significantly smaller than $0.1~\mu$m.

\begin{figure}
\begin{center}
\includegraphics[width=0.8\linewidth]{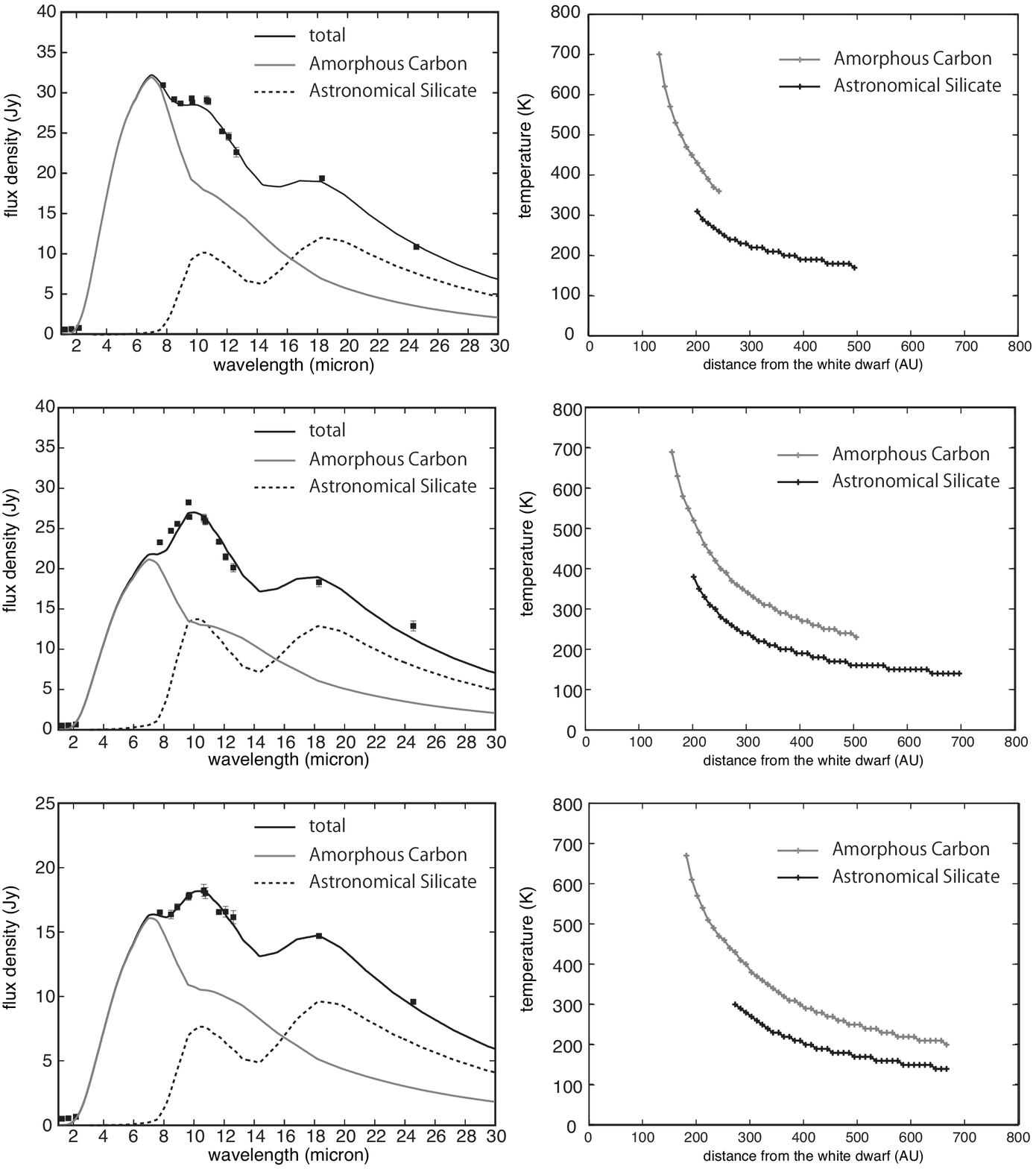}
\caption{The same as Fig.~\ref{fig_best_fit_SED_RT_1} but for $a_{a.car.}=0.01~\mu$m and $a_{a.sil.}=0.1~\mu$m.
 \label{fig_best_fit_SED_RT_3}
  }
\end{center}
\end{figure}

\begin{figure}
\begin{center}
\includegraphics[width=0.8\linewidth]{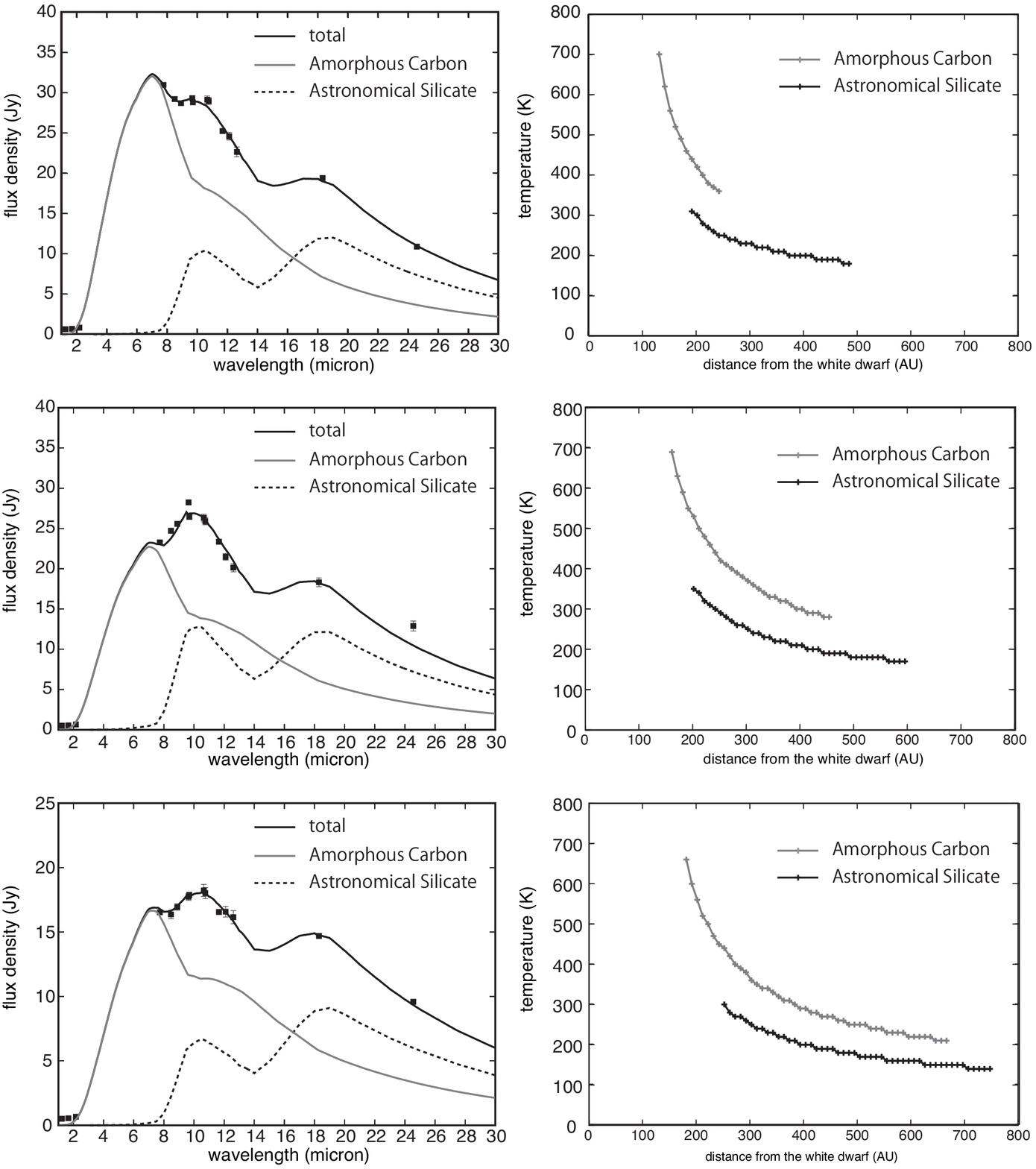}
\caption{The same as Fig.~\ref{fig_best_fit_SED_RT_1} but for $a_{a.car.}=0.01~\mu$m and $a_{a.sil.}=0.2~\mu$m.
 \label{fig_best_fit_SED_RT_4}
  }
\end{center}
\end{figure}

\begin{figure}
\begin{center}
\includegraphics[width=0.8\linewidth]{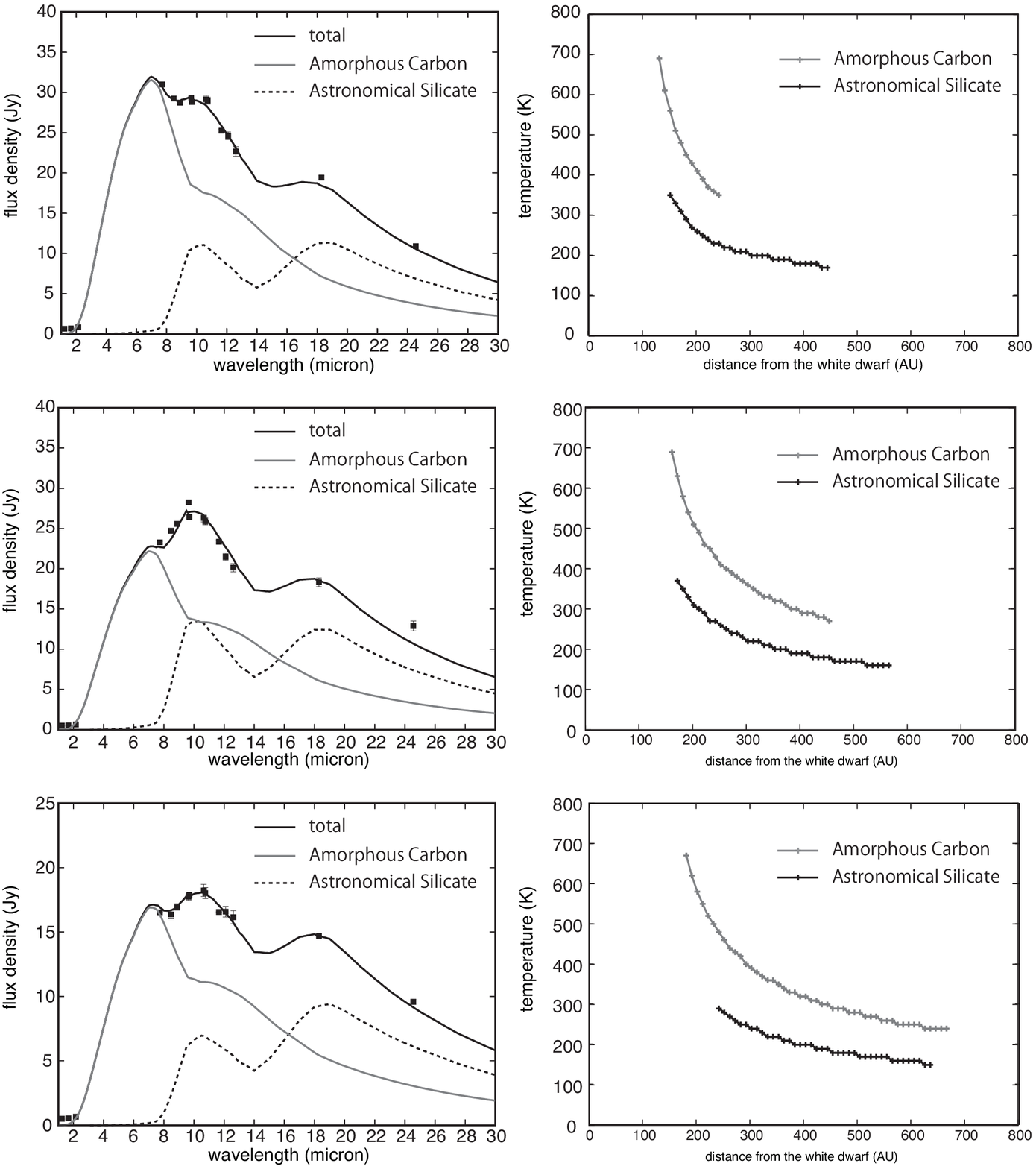}
\caption{The same as Fig.~\ref{fig_best_fit_SED_RT_1} but for $a_{a.car.}=0.01~\mu$m and $a_{a.sil.}=0.3~\mu$m.
 \label{fig_best_fit_SED_RT_5}
  }
\end{center}
\end{figure}

\begin{figure}
\begin{center}
\includegraphics[width=0.8\linewidth]{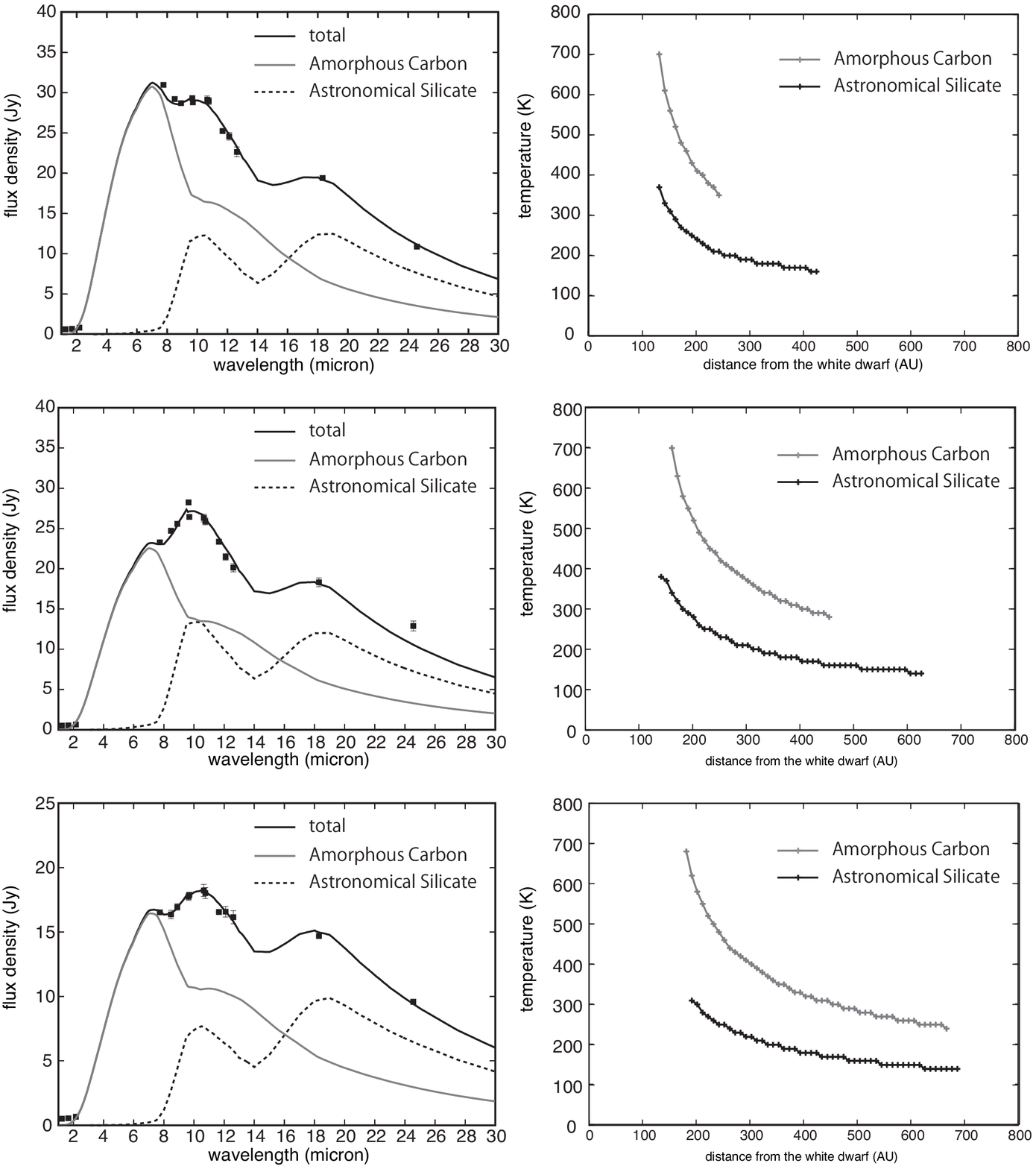}
\caption{The same as Fig.~\ref{fig_best_fit_SED_RT_1} but for $a_{a.car.}=0.01~\mu$m and $a_{a.sil.}=0.5~\mu$m.
 \label{fig_best_fit_SED_RT_6}
  }
\end{center}
\end{figure}

\begin{figure}
\begin{center}
\includegraphics[width=0.8\linewidth]{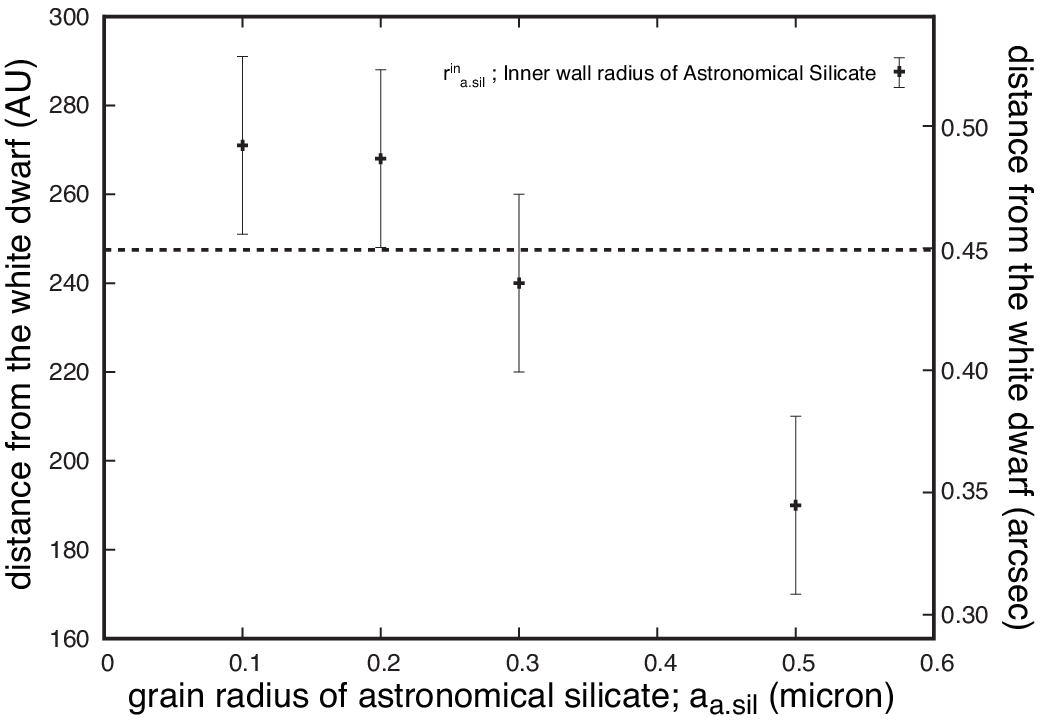}
\caption{The plot of the inner wall radius of astronomical silicate on Day 1947 obtained for each grain radius of $a_{a.sil.}=0.1, 0.2, 0.3, 0.5$~$\mu$m. The broken line indicates the semi-major axis length of the decomvolved Qb band image of V1280 Sco modeled with an elliptical Gaussian profile on Day 1947, i.e. 0".45.  \label{R_IN_AS}}
\end{center}
\end{figure}

The best-fit parameters obtained on Days 1272, 1616 and 1947 calculated for the cases of $a_{a.car.}=0.01~\mu$m, $a_{a.sil.}=0.1~\mu$m, $0.2~\mu$m, $0.3~\mu$m, and $0.5~\mu$m are summarized in Tables~\ref{tbl_parameters_geometry_3}, \ref{tbl_parameters_geometry_4}, \ref{tbl_parameters_geometry_5}, \ref{tbl_parameters_geometry_6}, respectively. The model spectra calculated with those best-fit parameters are shown in the left panels of Figs~\ref{fig_best_fit_SED_RT_3} -- ~\ref{fig_best_fit_SED_RT_6}. Temperatures of amorphous carbon and astronomical silicate are shown as a function of distance from the white dwarf in the right panels of Figs~\ref{fig_best_fit_SED_RT_3}-- ~\ref{fig_best_fit_SED_RT_6}. The best-fit $r^{in}_{a.sil.}$'s on Day 1947 are shown in Fig.~\ref{R_IN_AS} for different $a_{a.sil.}$ used in the SED analysis. Because $r^{in}_{a.sil.}$ must be smaller than the semi-major axis of the deconvolved Qb band image of V1280 Sco on Day 1947 (i.e. 0".45), we conclude that the typical size of silicate dust must be larger than 0.3~$\mu$m (see Fig.~\ref{R_IN_AS}). Moreover, the semi-major axis of the deconvolved emitting region measured in the Qb band, which is dominated by astronomical silicate, is systematically larger than that measured in the Si-1 band, which is dominated by amorphous carbon, on Days 1272, 1616 and 1947. This implies astronomical silicate is located farther from the white dwarf than amorphous carbon. We have found that $r^{in}_{a.sil.}$ is always larger than $r^{in}_{a.car.}$ when $a_{a.sil.}<0.5~\mu$m ($a_{a.car.}=0.01\mu$m; see Tables~\ref{tbl_parameters_geometry_3} -- ~\ref{tbl_parameters_geometry_6}). We therefore conclude the typical grain size of amorphous carbon as $a_{a.car.}=0.01~\mu$m and that of astronomical silicate as $0.3~\mu$m$<a_{a.sil.}<0.5~\mu$m. A relatively large grain size found for silicate dust in our analyses is consistent with the general understanding that the grain size in dust shells of novae is substantially larger than that in the interstellar medium \citep{geh98, wit01}.

\begin{table}[htbp!]
\begin{center}
\caption{
The best-fit parameters obtained for $a_{a.car.}=0.01~\mu$m and $a_{a.sil.}=0.1~\mu$m \label{tbl_parameters_geometry_3}
}
\scriptsize
\begin{tabular}{lccc}
\tableline \tableline
Epoch (days)            
&      1272        
&      1616         
&      1947        \\ \tableline
opening angle; $\phi$ [deg] 
& 24.0$\pm$0.5    
& 18.0$\pm$0.5    
& 15.0$\pm$0.5   \\
Amor. Car. initial number density; $n^{ini}_{a.car.}$ [cm$^{-3}$] 
& 2.0$^{+0.2}_{-0.2}$$\times$$10^{-3}$ 
& 1.3$^{+0.2}_{-0.2}$$\times$$10^{-3}$ 
& 1.1$^{+0.2}_{-0.1}$$\times$$10^{-3}$ \\
Amor. Car. inner radius; $r^{in}_{a.car.}$ [AU] 
& 1.3$^{+0.1}_{-0.1}$$\times$$10^{2}$ 
& 1.5$^{+0.1}_{-0.1}$$\times$$10^{2}$
& 1.7$^{+0.1}_{-0.2}$$\times$$10^{2}$ \\
Amor. Car. outer radius; $r^{out}_{a.car.}$ [AU] 
& 2.5$^{+0.3}_{-0.2}$$\times$$10^{2}$ 
& 4.6$^{+0.5}_{-0.5}$$\times$$10^{2}$ 
& 6.6$^{+0.6}_{-0.6}$$\times$$10^{2}$ \\

Astro. Sil. initial number density; $n^{ini}_{a.sil.}$  [cm$^{-3}$] 
& 1.3$^{+0.1}_{-0.1}$$\times$$10^{-6}$ 
& 1.8$^{+0.2}_{-0.2}$$\times$$10^{-6}$ 
& 1.5$^{+0.2}_{-0.2}$$\times$$10^{-6}$ \\
Astro. Sil. inner radius; $r^{in}_{a.sil.}$ [AU]
& 2.0$^{+0.2}_{-0.2}$$\times$$10^{2}$
& 2.0$^{+0.2}_{-0.2}$$\times$$10^{2}$
& 2.7$^{+0.2}_{-0.2}$$\times$$10^{2}$\\
Astro. Sil. outer radius; $r^{out}_{a.sil.}$ [AU]
& 5.0$^{+1.0}_{-0.5}$$\times$$10^{2}$
& 6.0$^{+1.0}_{-0.5}$$\times$$10^{2}$
& 6.7$^{+1.0}_{-0.5}$$\times$$10^{2}$\\
\tableline
\end{tabular}
\end{center}
\end{table}

\begin{table}[htbp!]
\begin{center}
\caption{
The best-fit parameters obtained for $a_{a.car.}=0.01~\mu$m and $a_{a.sil.}=0.2~\mu$m \label{tbl_parameters_geometry_4}
}
\scriptsize
\begin{tabular}{lccc}
\tableline \tableline
Epoch (days)            
&      1272        
&      1616         
&      1947        \\ \tableline
opening angle; $\phi$ [deg] 
& 24.0$\pm$0.5    
& 18.0$\pm$0.5    
& 15.0$\pm$0.5   \\
Amor. Car. initial number density; $n^{ini}_{a.car.}$ [cm$^{-3}$] 
& 2.1$^{+0.2}_{-0.2}$$\times$$10^{-3}$ 
& 1.2$^{+0.2}_{-0.2}$$\times$$10^{-3}$ 
& 1.3$^{+0.2}_{-0.2}$$\times$$10^{-3}$ \\
Amor. Car. inner radius; $r^{in}_{a.car.}$ [AU] 
& 1.3$^{+0.1}_{-0.1}$$\times$$10^{2}$ 
& 1.5$^{+0.1}_{-0.1}$$\times$$10^{2}$
& 1.7$^{+0.1}_{-0.2}$$\times$$10^{2}$ \\
Amor. Car. outer radius; $r^{out}_{a.car.}$ [AU] 
& 2.5$^{+0.3}_{-0.2}$$\times$$10^{2}$ 
& 4.6$^{+0.5}_{-0.5}$$\times$$10^{2}$ 
& 6.6$^{+0.6}_{-0.6}$$\times$$10^{2}$ \\

Astro. Sil. initial number density; $n^{ini}_{a.sil.}$  [cm$^{-3}$] 
& 1.6$^{+0.1}_{-0.1}$$\times$$10^{-7}$ 
& 1.8$^{+0.2}_{-0.2}$$\times$$10^{-7}$ 
& 1.8$^{+0.2}_{-0.2}$$\times$$10^{-7}$ \\

Astro. Sil. inner radius; $r^{in}_{a.sil.}$ [AU]
& 2.0$^{+0.2}_{-0.2}$$\times$$10^{2}$
& 2.0$^{+0.2}_{-0.2}$$\times$$10^{2}$
& 2.7$^{+0.2}_{-0.2}$$\times$$10^{2}$\\
Astro. Sil. outer radius; $r^{out}_{a.sil.}$ [AU]
& 4.9$^{+1.0}_{-0.5}$$\times$$10^{2}$
& 6.0$^{+2.0}_{-1.0}$$\times$$10^{2}$
& 7.0$^{+1.0}_{-0.5}$$\times$$10^{2}$\\
\tableline
\end{tabular}
\end{center}
\end{table}

\begin{table}[htbp!]
\begin{center}
\caption{
The best-fit parameters obtained for $a_{a.car.}=0.01~\mu$m and $a_{a.sil.}=0.3~\mu$m \label{tbl_parameters_geometry_5}
}
\scriptsize
\begin{tabular}{lccc}
\tableline \tableline
Epoch (days)            
&      1272        
&      1616         
&      1947        \\ \tableline
opening angle; $\phi$ [deg] 
& 24.0$\pm$0.5    
& 18.0$\pm$0.5    
& 15.0$\pm$0.5   \\
Amor. Car. initial number density; $n^{ini}_{a.car.}$ [cm$^{-3}$] 
& 2.2$^{+0.3}_{-0.2}$$\times$$10^{-3}$ 
& 1.3$^{+0.2}_{-0.2}$$\times$$10^{-3}$ 
& 1.0$^{+0.2}_{-0.2}$$\times$$10^{-3}$ \\
Amor. Car. inner radius; $r^{in}_{a.car.}$ [AU] 
& 1.3$^{+0.1}_{-0.1}$$\times$$10^{2}$ 
& 1.5$^{+0.1}_{-0.1}$$\times$$10^{2}$
& 1.7$^{+0.1}_{-0.2}$$\times$$10^{2}$ \\
Amor. Car. outer radius; $r^{out}_{a.car.}$ [AU] 
& 2.5$^{+0.3}_{-0.2}$$\times$$10^{2}$ 
& 4.6$^{+0.5}_{-0.5}$$\times$$10^{2}$ 
& 6.6$^{+0.6}_{-0.6}$$\times$$10^{2}$ \\

Astro. Sil. initial number density; $n^{ini}_{a.sil.}$  [cm$^{-3}$] 
& 7.1$^{+0.7}_{-0.7}$$\times$$10^{-8}$ 
& 7.9$^{+0.8}_{-0.8}$$\times$$10^{-8}$ 
& 6.3$^{+0.7}_{-0.7}$$\times$$10^{-8}$ \\
Astro. Sil. inner radius; $r^{in}_{a.sil.}$ [AU]
& 1.5$^{+0.1}_{-0.1}$$\times$$10^{2}$
& 1.7$^{+0.1}_{-0.1}$$\times$$10^{2}$
& 2.4$^{+0.2}_{-0.2}$$\times$$10^{2}$\\
Astro. Sil. outer radius; $r^{out}_{a.sil.}$ [AU]
& 4.5$^{+1.0}_{-0.5}$$\times$$10^{2}$
& 5.7$^{+2.0}_{-1.0}$$\times$$10^{2}$
& 6.4$^{+1.0}_{-0.5}$$\times$$10^{2}$\\
\tableline
\end{tabular}
\end{center}
\end{table}

\begin{table}[htbp!]
\begin{center}
\caption{
The best-fit parameters obtained for $a_{a.car.}=0.01~\mu$m and $a_{a.sil.}=0.5~\mu$m \label{tbl_parameters_geometry_6}
}
\scriptsize
\begin{tabular}{lccc}
\tableline \tableline
Epoch (days)            
&      1272        
&      1616         
&      1947        \\ \tableline
opening angle; $\phi$ [deg] 
& 24.0$\pm$0.5    
& 18.0$\pm$0.5    
& 15.0$\pm$0.5   \\
Amor. Car. initial number density; $n^{ini}_{a.car.}$ [cm$^{-3}$] 
& 2.1$^{+0.2}_{-0.2}$$\times$$10^{-3}$ 
& 1.2$^{+0.1}_{-0.1}$$\times$$10^{-3}$ 
& 0.9$^{+0.1}_{-0.1}$$\times$$10^{-3}$ \\
Amor. Car. inner radius; $r^{in}_{a.car.}$ [AU] 
& 1.3$^{+0.1}_{-0.1}$$\times$$10^{2}$ 
& 1.5$^{+0.1}_{-0.1}$$\times$$10^{2}$
& 1.7$^{+0.1}_{-0.2}$$\times$$10^{2}$ \\
Amor. Car. outer radius; $r^{out}_{a.car.}$ [AU] 
& 2.5$^{+0.3}_{-0.2}$$\times$$10^{2}$ 
& 4.6$^{+0.5}_{-0.5}$$\times$$10^{2}$ 
& 6.6$^{+0.6}_{-0.6}$$\times$$10^{2}$ \\

Astro. Sil. initial number density; $n^{ini}_{a.sil.}$  [cm$^{-3}$] 
& 7.1$^{+0.7}_{-0.7}$$\times$$10^{-8}$ 
& 7.9$^{+0.8}_{-0.8}$$\times$$10^{-8}$ 
& 6.3$^{+0.6}_{-0.6}$$\times$$10^{-8}$ \\
Astro. Sil. inner radius; $r^{in}_{a.sil.}$ [AU]
& 1.3$^{+0.1}_{-0.1}$$\times$$10^{2}$
& 1.3$^{+0.1}_{-0.1}$$\times$$10^{2}$
& 1.9$^{+0.2}_{-0.2}$$\times$$10^{2}$\\
Astro. Sil. outer radius; $r^{out}_{a.sil.}$ [AU]
& 4.3$^{+1.0}_{-0.5}$$\times$$10^{2}$
& 6.3$^{+2.0}_{-1.0}$$\times$$10^{2}$
& 6.9$^{+1.0}_{-0.5}$$\times$$10^{2}$\\
\tableline
\end{tabular}
\end{center}
\end{table}

The dust masses of amorphous carbon and astronomical silicate calculated for $a_{a.car.}=0.01~\mu$m and $a_{a.sil.}=0.3~\mu$m are given in Table~\ref{tbl_parameters_M}, where $\rho_{a.car.}$ (1.87 g cm$^{-3}$) and $\rho_{a.sil.}$ (3.3 g cm$^{-3}$) are adopted as the the densities of an amorphous carbon ($a.car.$) and astronomical silicate ($a.sil.$) dust particle, respectively. These values suggest that no significant dust mass evolution occurred from Day 1272 through 1947. Thus, the mass of amorphous carbon and astronomical silicate around V1280 Sco is $\sim$8$\times$10$^{-8}$M$_{\odot}$ and $\sim$4$\times$10$^{-7}$M$_{\odot}$, respectively.

\begin{table}[htbp!]
\begin{center}
\caption{The Dust Mass Evolution among Days 1272, 1616 and 1947. \label{tbl_parameters_M}
}
\scriptsize
\begin{tabular}{lcc}
\tableline \tableline
 Epoch (days)   & Amor. Carbon ($M_{\odot}$)  & Astro. Silicate ($M_{\odot}$)  \\
\tableline
 1272 & 6.6$^{+2.0}_{-1.4}$$\times$10$^{-8}$ & 3.4$^{+0.6}_{-0.4}$$\times$10$^{-7}$ \\ 
 1616 & 8.7$^{+1.7}_{-1.5}$$\times$10$^{-8}$ & 4.3$^{+1.3}_{-1.2}$$\times$10$^{-7}$ \\
 1947 & 8.4$^{+1.0}_{-1.7}$$\times$10$^{-8}$ & 4.0$^{+0.8}_{-0.8}$$\times$10$^{-7}$  \\
\tableline
\end{tabular}
\end{center}
\end{table}

\subsection{Dust Formation History around V1280 Sco{\label{DUST_HISTORY}}}
The results of our analyses have shown that the infrared SED evolution of V1280 Sco on Days 1272, 1616, and 1947 after the outburst is well reproduced by emission from a mixture of amorphous carbon and astronomical silicate that travel away from the white dwarf without any apparent mass evolution at such later epochs. Although the absolute values depend on the geometry, our dust model described in \S~\ref{SED_RT} has found the typical size of amorphous carbon dust as $a_{a.car.}=$0.01$~\mu$m and its mass as $M_{a.car.} \sim 8\times$10$^{-8}$M$_{\odot}$, and the corresponding values for astronomical silicate as $0.3 \mu$m $<a_{a.sil}<0.5 \mu$m and $M_{a.sil.} \sim 4 \times$10$^{-7}$M$_{\odot}$. We note that the masses of amorphous carbon dust obtained on Days 1272, 1616, and 1947 are much smaller than that of optically thick amorphous carbon on Day 150, i.e. 2.7$\times$10$^{-6}$M$_{\odot}$ (see \S~\ref{SED_D150}). Moreover, the typical grain size of amorphous carbon dust, i.e. $a_{a.car.}=$0.01$~\mu$m, obtained on Days 1272, 1616, and 1947 is much smaller than that of carbon grains that have formed in other CO novae \citep[e.g., $\sim$0.3~$\mu$m for Nova Serpentis 1978 on Day 70; ][]{geh80}. We suggest, therefore, that sputtering as a result of the interaction with the circumstellar medium as well as evaporation may have occurred for amorphous carbons during the period from Days 150 to 1272.

All the images taken on Day 1947, at shorter wavelengths (e.g., Si-1) which are dominated by amorphous carbon emission, or at longer wavelengths (e.g., Qa and Qb) which are dominated by silicate dust, exhibit an elongated structure along PA $= 20$ deg. The radius of the carbon emitting region was 0".40$\pm$0".03 and that of the silicate emitting region was 0".45$\pm$0".04 on Day 1947. Assuming that the nova ejecta travel with a constant velocity $v$, the projected distance $d$ to which the nova ejecta with a velocity range of 320 km s$^{-1}$ $< v <$ 620~km s$^{-1}$ \citep{nai12} can travel in 1947 days is 0".33$<d<$0".64, which clearly covers both amorphous carbon and silicate dust emitting regions observed on Day 1947. In addition, no infrared echoes from the pre-existing silicate dust were detected in our mid-infrared spectrum taken on Day $150$. It is therefore more likely that the silicate emission observed on Days 1272, 1616 and 1947 is produced either by newly formed dust grains in the expanding nova ejecta or as a result of interaction between the ejecta and the circumstellar medium, rather than by pre-existing circumstellar dust grains.

The dust formation scenario around V1280 Sco suggested by our analyses is that the amorphous carbon dust is formed in the nova ejecta followed by the silicate dust formation. Such dual dust chemistry has been reported in a number of other novae such as QV Vul \citep{geh92} and V705 Cas \citep{mas98,eva97}. The dust formation sequence in V1280 Sco is consistent with those novae, where the carbon-rich dust is identified early in the outburst and oxygen-rich dust later in the outburst \citep{geh98}. Such chemical multiplicity of the dust formation requires a chemical gradient and/or heterogeneity in the nova ejecta. In the case of V1280 Sco, such conditions may have been produced by multiple ejection events of nova winds that have been observed, at least, on Days 10.5 and $\sim$110 \citep{che08} and that may be composed of different materials dredged up from the underlying CO white dwarf core. 

Another possibility is that dust is formed not just in the nova ejecta but also in the pre-existing circumstellar medium as a result of its interaction with shocks generated by the ejecta. If the chemical properties in the nova ejecta and the pre-existing circumstellar medium were quite different, the amorphous carbon dust is formed in the carbon-rich nova ejecta and the silicate dust in the oxygen-rich circumstellar medium. The presence of many cool clumpy gas clouds produced via interactions between the pre-existing cool circumstellar gas and high velocity gas ejected in the nova explosion has been proposed by \citet{sad10}. The formation of silicate dust could have occurred in those cool clumpy gas clouds if the cool circumstellar material is oxygen-rich. 

Optical and ultraviolet studies, however, indicate that the composition of matter ejected by classical novae is typically oxygen-rich \citep{geh98,sta98}. If this is also the case in V1280 Sco, the formation mechanism of amorphous carbon dust has to be carefully addressed. \citet{cla99} suggest that carbon dust can form even in oxygen-rich (C/O$<1$) wind if free carbon exists abundantly and if carbon nucleation occurs despite rapid oxidation. If this is the case in the nova ejecta of V1280 Sco, the amorphous carbon dust may have been produced in a region closer to the white dwarf, where CO formation is disturbed by harsh radiation and/or by Compton electrons, than the CO forming region. Then, the optically thick amorphous carbon dust may have blocked the ultraviolet radiation from the WD, thus providing an environment for the ejecta gas, which no longer contains abundant free carbon atoms behind the optically thick amorphous carbon dust, to cool quickly. This situation may have led silicate dust to condense in the gas at a relatively high density, which resulted in attaining relatively large grain size of $0.3~\mu$m$<a_{a.sil.}<0.5~\mu$m.


\section{Spectral Features over the Continuum Emission \label{sect6}}
\begin{figure}
\begin{center}
\includegraphics[width=1.0\linewidth]{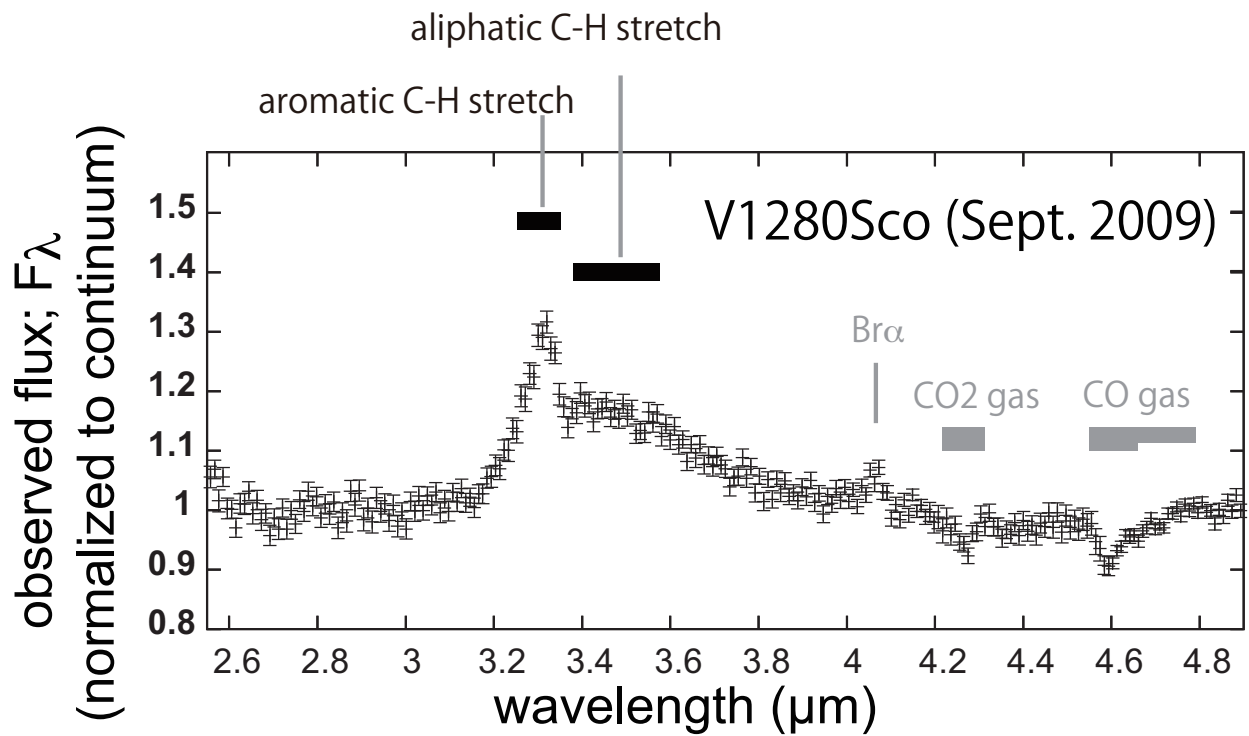}
\caption{The near infrared spectrum of V1280 Sco on Day 940 observed with AKARI/IRC. The spectrum is normalized to the pseudo continuum modeled by 2nd-order polynomial function through the continuum points at $\sim$3.0$\mu$m, $\sim$4.1$\mu$m and $\sim$4.8$\mu$m. The wavelength ranges of features arising from the aromatic C-H stretches and aliphatic C-H stretches are indicated with black horizontal bars. The wavelength ranges of absorption features possibly carried by CO$_2$ gas at 4.26$\mu$m and CO gas at 4.6$\mu$m are shown with gray horizontal bars. Small feature at 4.05$\mu$m is attributed to Br$\alpha$.
\label{akari_NG}
}
\end{center}
\end{figure}

Notable in the near infrared spectrum of V1280 Sco on Day 940 obtained with the AKARI infrared camera (IRC) is the presence of the UIR band at 3.3$~\mu$m (see Figure~\ref{akari_NG}). The shape of this feature is characterized by unusually strong broad red wings at 3.4--3.6$~\mu$m. The prominent 3.4--3.6$~\mu$m wing in the near infrared spectrum of V1280 Sco on Day 940 is consistent with the spectral characteristics of the UIR bands observed in other novae such as V705 Cas \citep{eva97} and V842 Cen \citep{hyl89}. The 3.3$~\mu$m feature is usually attributed to an aromatic C-H stretching mode, while the 3.4--3.6$~\mu$m wing to aliphatic C-H stretching modes \citep{all89}. We therefore suggest that a higher aliphatic-to-aromatic ratio exists in the carrier of the UIR bands in V1280 Sco on Day 940 than in other Galactic objects. As discussed by \citet{eva97}, hydrogenated amorphous carbons \citep[HACs;][]{bor87,dul88} rather than polycyclic aromatic hydrocarbons (PAHs) could be a plausible candidate for the carrier of UIR bands observed in V1280 Sco. Alternatively, overabundance of nitrogen in the nova wind may also be responsible for the relatively stronger 3.4--3.6$~\mu$m to 3.3$~\mu$m ratio \citep{and90,roc96}. 

The near infrared spectrum of V1280 Sco on Day 940 also shows the absorption features due to CO gas at 4.6$~\mu$m and, possibly, to CO$_2$ gas at 4.26$~\mu$m. The continuum emission at this wavelength is predominantly produced by amorphous carbon as indicated by our SED analyses described in \S~\ref{SED_RT}. The presence of CO gas in absorption may support the dust formation scenario discussed in \S~\ref{DUST_HISTORY}, that the amorphous carbon dust has been formed in a region closer to the white dwarf than the CO forming region. 


\begin{figure}
\begin{center}
\includegraphics[width=0.6\linewidth]{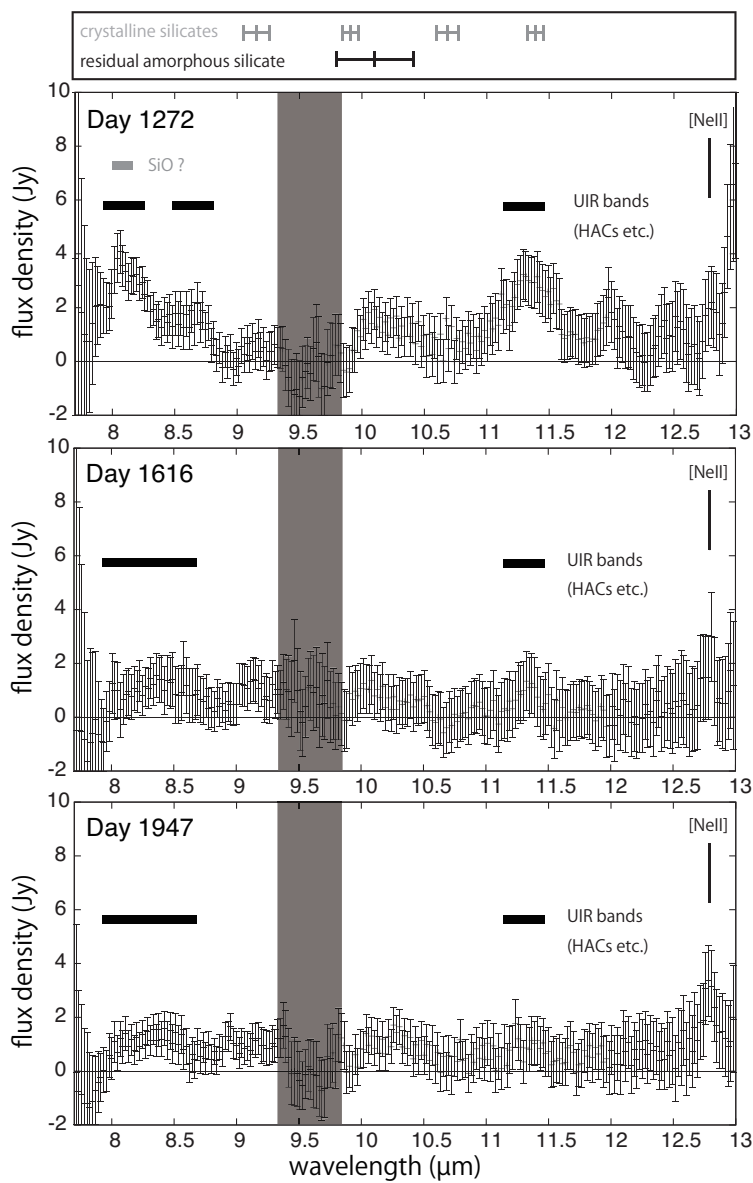}
\caption{The mid infrared spectra of V1280 Sco on Days 1272, 1616 and 1947 observed with Gemini/T-ReCS. The best-fit model spectrum obtained as a result of the SED fitting (see section~\ref{SED_RT}) at each epoch is used to define the continuum and is subtracted. The wavelength ranges of the UIR band features recognized in our observations are shown with black horizontal bars. Peak positions of 9.2, 9.8, 10.7 and 11.4$~\mu$m features of crystalline silicates (e.g., silica, forsterite, enstatite and diopside) reported in \citet{mol02} are also shown with gray horizontal errorbars as a reference.
\label{trecs_NL}
}
\end{center}
\end{figure}

Figure~\ref{trecs_NL} shows the continuum-subtracted N-band low-resolution spectra of V1280 Sco on Days 1272, 1616 and 1947 obtained with Gemini-S/TReCS. At Day 1272, the spectrum is characterized by the presence of emission features at 8.1~$\mu$m and 11.3~$\mu$m. Silicon monoxide (SiO) is one of the candidates for the carrier of the 8.1~$\mu$m feature. The 8.1~$\mu$m feature recognized on Day 1272 has become significantly weaker by Day 1616. This can be interpreted as a decrease in the temperature of ejecta gas where SiO molecules are no longer excited. SiO molecules are expected to be abundant in the chemical pathway leading to the silicate dust formation \citep{koz89}. There exists observational evidence for an on-going silicate dust formation in the ejecta of Type II-P supernova 2004et for the period of days 300--464 accompanied by the mass decline of SiO \citep{kot09}. In the case of V1280 Sco, however, while the decline of 8.1$~\mu$m feature has been observed between Days 1272 and 1616, it was not apparent if any significant mass evolution of silicate dust had taken place on Days 1272, 1616 and 1947 (see \S~\ref{SED_RT}). This fact will not support the interpretation that SiO is the carrier of the 8.1$~\mu$m feature. 

Alternatively, taking into account the similarity between the {\it{class C}} UIR bands \citep{pee02} and the emission features at 8.1~$\mu$m and 11.3~$\mu$m in the spectrum of V1280 Sco on Day 1272, HACs \citep{bor87,dul88,slo07} and/or mixed aromatic-aliphatic organic nanoparticles \citep{kwo11} can be responsible for the emission features at 8.1~$\mu$m and 11.3~$\mu$m. These interpretations are consistent with the results obtained from the AKARI/IRC spectroscopy on Day 940. 

Finally, although astronomical silicate is required to reproduce the continuum emission, none of crystalline silicate features \citep[e.g., 9.2~$\mu$m for silica, 9.8~$\mu$m for forsterite and enstatite, 10.7~$\mu$m for enstatite, and 11.4~$\mu$m for forsterite and diopside; ][]{mol02} was detected in the continuum-subtracted N-band spectra of V1280 Sco at any epochs. This suggests that nucleation of silicate around V1280 Sco may have happened at relatively low temperatures \citep[e.g., $<800$~K; ][]{mol05} and that any further crystallization processes such as the annealing by UV photons \citep{nut90} or partial crystallization by electron irradiation \citep{car01} in the harsh environment around the white dwarf have not occurred for silicate dust around V1280 Sco at least on a timescale of 2000 days from the outburst.


\pagebreak

\section{Summary}
We present the results of the late time infrared multi-epoch observations of the dust forming nova V1280 Sco. The temperature and mass of dust formed in the nova wind are examined based on the spectral decomposition of the infrared SED obtained at each epoch. The SED on Day 150 suggest the presence of optically thick amorphous carbon dust of mass $M_{a.car.} \sim 2.7\times 10^{-6}$M$_{\odot}$. The SED evolution in a period from Days $1272$ to $1947$ can be explained by emission from amorphous carbon dust of mass $M_{a.car.} \sim 8\times 10^{-8}$M$_{\odot}$ with the representative size of $a_{a.car}=$0.01~$\mu$m, combined with that from astronomical silicate of mass $M_{a.sil.} \sim 4\times 10^{-7}$M$_{\odot}$ with the representative size of 0.3~$\mu$m$<a_{a.sil.}<0.5$~$\mu$m. They both traveled farther away from the white dwarf without apparent mass evolution from Days $1272$ to $1947$. The decrease in the mass of amorphous carbon in the period from Days 150 to 1272 and the extraordinary small grain size of amorphous carbon ($a_{a.car}=$0.01~$\mu$m) on Days 1272, 1616 and 1947 compared with other CO novae may suggest that sputtering and/or evaporation may have occurred for amorphous carbons during the period from Days 150 to 1272.

The dust formation scenario that emerges from our observations is that amorphous carbon dust is initially produced in a region closer to the white dwarf, which then blocks the ultra violet radiation from the WD, to provide an environment for the O-rich gas to cool quickly and to condense into silicate dust from the gas at a relatively high density. Cool clumpy clouds produced via interactions between the pre-existing circumstellar gas and high velocity nova ejecta may have provided a favorable condition for the silicate dust to attain a large grain size of $0.3~\mu$m$<a_{a.sil.}<0.5$~$\mu$m.

The near- to mid-infrared spectra obtained on Days $940$ and $1272$ have exhibited signs of the UIR emission bands which are more plausibly expected to be carried by hydrogenated amorphous carbons and/or mixed aromatic-aliphatic organic nanoparticles rather than PAHs. The absence of those features in the mid-infrared spectra of V1280 Sco on Days $1616$ and $1947$ suggests that the carriers of the UIR emission bands observed on Day $1272$ cannot survive the harsh circumstellar environment provided by the white dwarf almost reaching the nebular phase.

Further observations of dust forming novae with sufficient time resolution and monitoring duration are indispensable to understand how the dusty environment characterized both by carbon-rich and oxygen-rich chemistry can be achieved around novae. Mid-infrared studies with extremely high spatial resolution ($\sim$0".1) with an integral field unit (IFU), which will be achieved by the mid-infrared instrument with AO system onboard 30m class telescopes \citep[e.g., TMT/MICHI+MIRAO][]{pac12,chu06}, will play a crucial role to demonstrate the geometries of newly formed dust in the nova wind and the pre-existing circumstellar dust. In terms of the time resolution, in particular during the silicate dust formation phase, TAO/MIMIZUKU \citep{kam14} will play an important role to determine the physical parameters that control the nucleation of silicate dust. A wider wavelength coverage of SOFIA/FORCAST \citep{ada12} will be important to investigate the dust properties (crystallization and/or amorphization) of silicates from the mineralogical point of view. Finally, the spectroscopic capability of SPICA \citep{nak14} both in mid- and far-infrared with the highest sensitivity ever \citep{kat15} will demonstrate the longer-term evolution of dust from the epoch of its formation in the circumstellar environment until it becomes a member of the interstellar dust.

\acknowledgments

This research is based in part on data collected at the Subaru Telescope, which is operated by the National Astronomical Observatory of Japan and those obtained at the Gemini Observatory via the time exchange program between Gemini and the Subaru Telescope. The authors are grateful to all the Subaru telescope staff members and Gemini South staff members, particularly, to J. Rodomzki and P. Gomez for their support during the observations with Gemini-S/T-ReCS. The Gemini Observatory is operated by the Association of Universities for Research in Astronomy, Inc., under a cooperative agreement with the NSF on behalf of the Gemini partnership: the National Science Foundation (United States), the National Research Council (Canada), CONICYT (Chile), the Australian Research Council (Australia), Minist\'{e}rio da Ci\^{e}ncia, Tecnologia e Inova\c{c}\~{a}o (Brazil) and Ministerio de Ciencia, Tecnología e Innovaci\'{o}n Productiva (Argentina). This research is based in part on observations with AKARI, a JAXA project with the participation of ESA. The authors thank all the members of the AKARI project for their continuous support. A part of the data in this paper was obtained by the Optical \& Near-Infrared Astronomy Inter-University Cooperation Program, supported by the MEXT of Japan. The authors would like to thank T. Yoshikawa, K. Tsumura, M. Shirahata for their help in performing the observation at IRSF. The authors are grateful to M. Kato and I. Hachisu for providing us the parameters of the white dwarf of V1280 Sco based on the theoretical calculations. The authors are grateful to M. Matsuura for providing us useful comments.




\end{document}